\documentclass[prl,twocolumn]{revtex4}

\usepackage{color, graphics, graphicx, epsfig, latexsym, amsmath, lmodern}

\graphicspath{{/Users/julien/Documents/Drosophila/Papers/Positional_Information_Tech/Figures}}

\begin{document}

\title{Positional information, positional error, and read-out precision in morphogenesis: a mathematical framework}
\author{Ga\v{s}per Tka\v{c}ik$^{a}\footnote{Corresponding author: gasper.tkacik@ist.ac.at}$, Julien O. Dubuis$^{b,c}$, Mariela D. Petkova$^b$, Thomas Gregor$^{b,c}$}
\affiliation{$^a$Institute of Science and Technology Austria, A-3400 Klosterneuburg, Austria\\
${}^b$Joseph Henry Laboratories of Physics and ${}^c$Lewis Sigler Institute for Integrative Genomics, Princeton University, New Jersey 08544, USA
}

\begin{abstract}
The concept of positional information is central to our understanding of how cells in a multicellular structure determine their developmental fates. Nevertheless, positional information has neither been defined mathematically nor quantified in a principled way. Here we provide an information-theoretic definition in the context of developmental gene expression patterns and examine which features of expression patterns increase or decrease positional information. We connect positional information with the concept of positional error and develop tools to directly measure information and error from experimental data. We illustrate our framework for the case of gap gene expression patterns in the early \emph{Drosophila} embryo and show how information that is distributed among only four genes is sufficient to determine developmental fates with single cell resolution. Our approach can be generalized to a variety of different model systems; procedures and examples are discussed in detail.
\end{abstract}

\maketitle
%
%
%
%%%% INTRODUCTION
%
%
%

{\footnotesize
\tableofcontents
}

\section{Introduction}

Central to the formation of multicellular organisms is the ability of cells with identical genetic material to acquire distinct cell fates according to their position in a developing tissue \cite{Lawrence92,Kirschner97}. While many mechanistic details remain unsolved, there is a  wide consensus that cells acquire knowledge about their location by measuring local concentrations of various form-generating molecules, called ``morphogens'' \cite{Turing52,Wolpert69}. In many cases, these morphogens are transcription factor proteins that control the activity of other genes, themselves coding for transcription factors, resulting in a regulatory network whose successive layers produce ever more refined spatial patterns of gene expression \cite{Dassow2000,Tomancak2007,Fakhouri:2010,Jaeger:2011}. The systematic variation in the concentrations of these morphogens with position defines a chemical coordinate system, used by  cells to determine their location \cite{Nusslein91,StJohnston92,Grossni94}. Morphogens are thus said to contain ``positional information'', which is processed by the genetic network, ultimately giving rise to cell fate assignments that are very reproducible across the embryos of the same species \cite{Wolpert2011}.

The concept of positional information has been widely used as a qualitative descriptor and has had an enormous success in shaping our current understanding of spatial patterning in developing organisms \cite{Wolpert69,Tickle1975,French1976,Driever1988b, Meinhardt:1988, Struhl1989, Reinitz:1995, Schier2005, Ashe2006, Jaeger:2006, Bokel2013, Witchley2013}. Mathematically, however, positional information has not been rigorously defined. Specific morphological features during early development have been studied in great detail and shown to occur reproducibly across wild-type embryos \cite{Gierer91, Gregor07a, Okabe2009, MSB, Liu2013}, while perturbations to the morphogen system resulted in systematic shifts of these same features \cite{Driever1988b,Struhl1989, Liu2013}. This established a causal---but not quantitative---link between the positional information encoded in the morphogens and the resulting body plan. Building on previous as well as on new results, we provide the missing quantitative link by proposing a mathematical formalism for positional information.

Specifically, we set out to achieve the following goals: first, we define positional information and positional error formally within the context of Shannon's information theory \cite{Shannon48}; second, we identify features of developmental gene expression that increase or decrease the information; and third, we establish the inference tools necessary to measure positional information in real datasets.  We focus on conceptual and data analysis details, with the explicit aim of presenting our framework such that it can be applied to different developmental systems. In two related papers, we have presented the experimental details, limitations, and interpretation when applying the proposed framework to the early \emph{Drosophila} embryo \cite{MSB,PNASPI}.

In \emph{Drosophila}, the entire body plan of the future adult organism is established by a hierarchical network of interacting genes during the first three hours of embryonic development \cite{NW1980,Akam1987,Ingham88,Spradling:1993,Papatsenko2009}. The hierarchy is composed of three layers: long-range protein gradients that span the entire long axis of the egg \cite{Driever:1988a}, gap genes expressed in broad bands \cite{Jaeger:2011}, and pair-rule genes that are expressed in a regular seven-striped pattern \cite{{Lawrence89}}. These genes typically encode transcription factors, hence allowing the layers to interact. Positional information is provided to the system solely via the first layer, which is established from maternally supplied and highly-localized mRNA that act as protein sources for the maternal gradients \cite{Nusslein91, StJohnston92, Anderson98, Ferrandon:1994, Little:2011}. The  network  then uses these inputs to generate distinct rows of nuclei that express the downstream genes in unique and distinguishable combinations in a process that takes two to three hours to complete \cite{Gergen:1986}. 

It is remarkable that such precision can be achieved in such a short amount of time using only a few handfuls of genes. Gene expression is subject to intrinsic fluctuations, which trace back to the randomness associated with regulatory  interactions between molecules present at low absolute copy numbers \cite{vankampen,tsimring}. Moreover, there is random variability not only within, but also between, embryos, for instance in the strength of the  morphogen sources \cite{bollenbach}. These biophysical limitations---e.g. in the number of signaling molecules, the time available for morphogen readout, and the reproducibility of initial and environmental conditions---place severe constraints on the ability of the developmental system to generate reproducible gene expression patterns \cite{Gregor07a}. 

Traditionally,  precision and reproducibility have been thought of as the ability to generate spatially sharp boundaries between broad bands of gene expression, where each gene would be either ``on'' or ``off'' in each of the expression domains. This view, which starts with the assumption that there are only two biologically meaningful levels of  expression, has been challenged recently by showing that the morphogen \emph{bicoid} transmits 1.5 bits of information to the gap gene \emph{hunchback}; this would suffice for three (instead of two) distinguishable levels of \emph{hunchback} response \cite{Tkacik08}. Moreover, a total of, e.g., four gap genes, each of which can  provide at most one bit (``on'' or ``off'') of information, is insufficient to uniquely specify even in principle the fates of more than $2^4$ nuclei, while the number of nuclei along the relevant segment of long axis of the \emph{Drosophila} embryo is closer to $\approx 2^6$ \cite{PNASPI}. Taken together, these considerations indicate that gap genes could be more than just ``binary switches,'' encoding a single bit of positional information each in their expression domains. To address these conflicting interpretations rigorously, we need to be able to make \emph{quantitative} statements about the positional information of spatial gene expression profiles without presupposing which features of the profile (e.g. sharpness of the boundary, size of the domains, position-dependent variability, etc) encode the information.

Here we  make the case that the relevant measure for positional information is the \emph{mutual information} $I$---a central information-theoretic quantity defined by Shannon in the 1940s---between expression profiles of the gap genes and position in the embryo. Briefly, for a given level of gene expression noise within an embryo and a given level of gene profile variability across embryos in a population, the quantity $I$  represents the binary logarithm of the number of rows along the embryo axis that have distinguishable gene expression levels. In other words, $2^I$ is an upper bound to the possible number of distinct equiprobable cellular identities. We show how positional information puts mathematical limits to the ability  with which cells in the developing \emph{Drosophila} embryo can infer their position if they respond to gap gene concentrations (and thus the morphogen gradient) alone. This  allows us to quantitatively decide whether the  picture of sharp ``on''/``off'' domains of gene expression is sufficient, and to ask how variability across embryos impedes the ability of the patterning system to transmit positional information. In order to apply this information-theoretic approach to real data, we solve a number of technical challenges related to the limitations of the current experimental setups. We report on these innovations in detail in order to prepare our approach for a straightforward generalization to other developmental systems.

\section{Results}

\subsection{Theoretical foundations}

In this  section we establish the information-theoretic framework for positional information carried by spatial patterns of gene expression. To develop an intuition, we start with a one-dimensional toy example of a single gene, which will be generalized later to a many-gene system. We present  scenarios where  positional information is stored in different qualitative features of gene expression patterns. To capture that intuition mathematically, we give a precise definition of positional information for one and for multiple genes. Finally, we show how a quantitative formulation of positional information is related to ``decoding,'' i.e. the ability of the nuclei to infer their position in the embryo.

\subsubsection{``Positional information'' in spatial gene expression profiles}

\begin{figure}
\centering
\includegraphics[width = 3.3in]{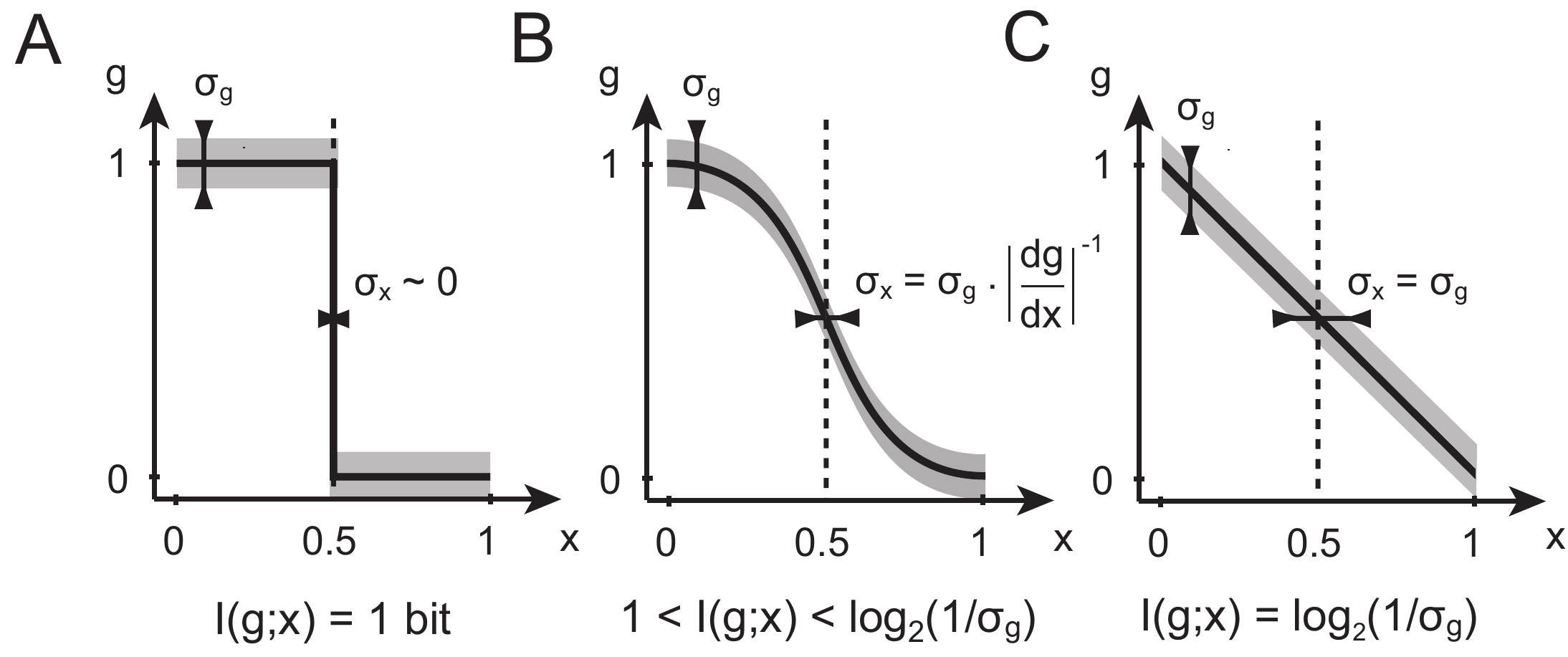}
\caption{{\bf Positional information encoded by a single gene.}  Shown are three hypothetical mean gene expression profiles $\bar{g}(x)$ as a function of position $x$, with spatially constant variability $\sigma_g$ (shaded area). {\bf A)} A step function  carries (at most) 1 bit of positional information, by perfectly distinguishing between ``off'' (not induced, posterior) and ``on'' (fully induced, anterior) states. {\bf B)} The boundary of the sigmoidal $g(x)$ function is now wider, but  the total amount of encoded positional information can be higher than 1 bit because the transition region itself is distinguishable from the ``on'' and ``off'' domains.  {\bf C)} A linear gradient has no well-defined boundary, but nevertheless provides a further increase in  information---if $\sigma_g$ is low enough---by being equally sensitive to position at every $x$. }
\label{InfoVSPrecision}
\end{figure}

Let us consider the simplest possible example where the expression of a single gene $G(x)$ varies with position $x$ along the axis of a one-dimensional embryo. We choose  units of length such that $x=0$ and $x=1$ correspond to the anterior and posterior poles of the embryo, respectively. Suppose we are able to quantitatively measure the profile of such a gene along the anterior-posterior axis in $\mathcal{N}$ embryos, labeled with an index $\mu=1,\dots,\mathcal{N}$. Such measurements of the light intensity profile of  fluorescently labelled antibodies against a particular gene product yield $G^{(\mu)}(x)$, where $G$ is the quantitative readout in embryo $\mu$ of the gene expression level. From a collection of embryos---after suitable data processing steps described later---we can then extract two statistics: the position-dependent ``mean profile,'' capturing the prototypical gene expression pattern, and the position-dependent variance across  embryos, which measures the degree of embryo-to-embryo variability or the reproducibility of the mean profile. We can transform the measurements  $G(x)$ into profiles $g(x)$ with rescaled units, such that the mean profile $\bar{g}(x)$ is normalized to 1 at the maximum and to 0 at the minimum along $x$. After these steps, our description of the system consists of the mean profile, $\bar{g}(x)$, and the variance in the profile, $\sigma_g^2(x)$. 

How much can a nucleus learn about its position if it expresses a gene at level $g$? We will compare three idealized cases, where we pick the shape of $\bar{g}(x)$ by hand and assume, for the start, that the variance is constant, $\sigma_g^2(x)=c$. The first case is illustrated in Fig.~\ref{InfoVSPrecision}A, where a step-like profile in $\bar{g}$ splits the embryo into two domains of gene expression:  an anterior ``on'' domain, where $\bar{g}(x) = 1$ for $x<x_0$, and an ``off'' domain in the posterior, $x>x_0$, where $\bar{g}(x)=0$. This arrangement has an extremely precise, indeed infinitely sharp, boundary at $x_0$; if we think that the precision of the boundary is the biologically relevant feature in this system, this arrangement would correspond to an ideal patterning gene. But how much information can nuclei extract from such a profile? If $x_0$ were $1/2$, the boundary would reproducibly split the embryo into two equal domains: based on reading out the expression of $g$, the nucleus could decide whether it is in the anterior or posterior, a binary choice that is equally likely prior to reading out $g$. As we will see, the positional information needed (and provided by such a sharp profile!) to make a clear two-way choice between two a priori equally likely possibilities equals 1 bit.

Can a profile of a different shape do better? Figure~\ref{InfoVSPrecision}B shows a somewhat more realistic sigmoidal shape that has a steep, but not infinitely sharp, transition region. If the variance is small enough, $\sigma_g^2\ll 1$, this profile can be more informative about the position. Nuclei far at the anterior still have $\bar{g}\approx 1$ (full induction or the ``on'' state), while nuclei at the posterior still have $\bar{g}\approx 0$ (the ``off'' state). But the graded response in the middle defines new expression levels in $g$ that are significantly different from both $0$ and $1$. A nucleus with $g\approx 0.5$ will thus ``know'' that it is neither in the anterior nor in the posterior. This system will therefore be able to provide more positional information than the sharp boundary which is limited by 1 bit. Clearly, this conclusion is valid only insofar as the variance $\sigma_g^2$ is low enough; if it gets too big, the intermediate levels of expression in the transition region can no longer be distinguished and we are back to the 1 bit case. 

The extreme contrast to the infinitely sharp gradient is the linear gradient, depicted in Fig.~\ref{InfoVSPrecision}C. Wolpert already proposed that linear gradients might be efficient in encoding positional information \cite{Wolpert69}, and indeed we can extend the argument for the sigmoidal case to convince ourselves that if $\sigma_g^2$ is not a function of position, the linear gradient is the best choice. Consider starting at the anterior and moving towards the posterior: as soon as we move far enough in $x$ that the change in $\bar{g}(x)$ is above $\sigma_g$, we have created one more distinguishable level of expression in $g$, and thus a group of nuclei that, by measuring $g$, can differentiate themselves from their anteriorly-positioned neighbors. Finally, this reasoning gives us a hint about how to generalize to the case where the variance $\sigma_g^2$ depends on position, $x$. What is important is to count, as $x$ covers the range from anterior to posterior, how much $\bar{g}(x)$ changes  \emph{in units of the local variability, $\sigma_g(x)$}---it will turn out that this is directly related to the mutual information between $g$ and position.

Thus a sharp and reproducible boundary can correspond to a profile that does not encode a lot of positional information, and a linear profile where the boundary is not even well-defined can encode a high amount  of positional information. Ultimately, whether or not there are intermediate distinguishable levels of gene expression depends on the variability in the profile. Therefore, any measure of positional information \emph{must} be a function of both $\bar{g}(x)$ and $\sigma_g^2(x)$.

\begin{figure}
\centering
\includegraphics[width = 3.3in]{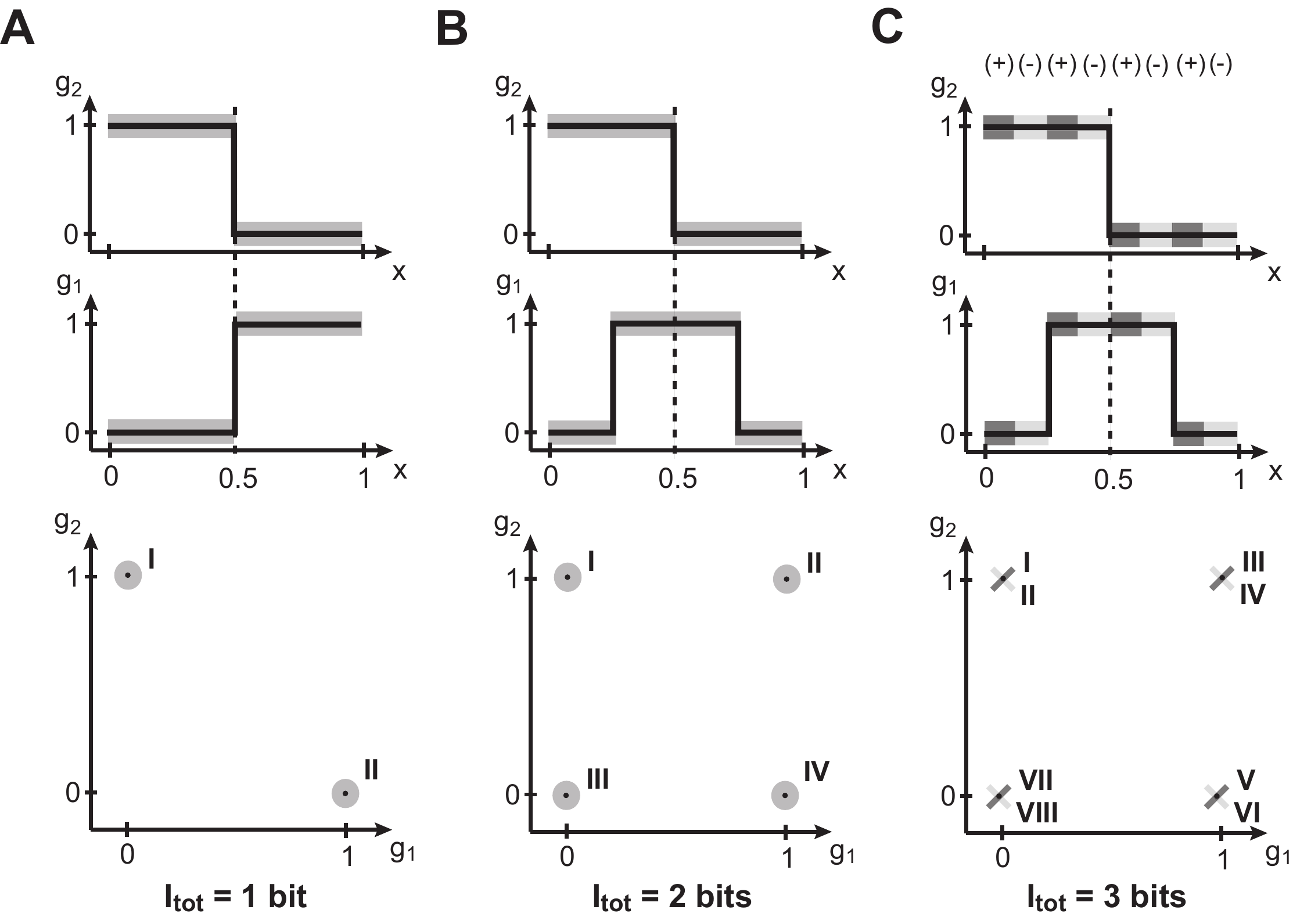}
\caption{{\bf Positional information encoded by two genes.} Spatial gene expression profiles are shown in the top row, while the bottom row schematically enumerates all distinguishable combinations of gene expression (roman numerals) across the embryo. {\bf A)} Genes $g_1, g_2$ are step functions, each encoding 1 bit of information. While the profiles are not the same, they are perfectly redundant, and the total number of jointly encoded ``states'' is only 2; this setup thus  conveys only one bit of positional information, the same as each gene alone. {\bf B)} The  mean profiles of $g_1,g_2$ have been displaced such that the redundancy is broken and the total information encoded is 2 bits (4 distinct states of joint gene expression). {\bf C)} If in addition to the mean profile shape the downstream layer can read out the (correlated) fluctuations of $g_1,g_2$, a further increase in information is possible. In this toy example, fluctuations are correlated $(+)$ and anti-correlated $(-)$ in various spatially separated regions, which allows the embryo to use this information along with the mean profile shape to distinguish 8 distinct regions, bringing the total encoded information to 3 bits.}
\label{ShapesANDCorr}
\end{figure}

Does the ability of the nuclei to infer their position automatically improve if they can simultaneously read out the expression levels of more than one gene? Figure~\ref{ShapesANDCorr}A shows the case where two genes, $g_1$ and $g_2$, do not provide any more information than each one of them provides separately, because they are completely redundant. Redundant does not mean equal---indeed, in Fig.~\ref{ShapesANDCorr}A the profiles are different at every $x$---but they are perfectly correlated (or dependent): knowing the expression level of $g_1$ one knows exactly the level of $g_2$, so $g_2$ cannot provide any additional new information about the position. In general, redundancy can help compensate for detrimental effects of noise when noise is significant, but this is not the case in the toy example at hand.

The situation is completely different if the two profiles are shifted relative to each other by, say, 25\% embryo length. Note that none of the individual profile properties have changed; both are still infinitely sharp with two states of gene expression, and half of the nuclei express in each state. However, the two genes now partition the embryo into 4 different segments: the anterior-most domain which is combinatorially encoded by the gene expression pattern $\bar{g}_1=\bar{g}_2=0$, the second domain with $\bar{g}_1=0,\;\bar{g}_2=1$, the third domain with $\bar{g}_1=\bar{g}_2=1$ and the last domain with $\bar{g}_1=1,\;\bar{g}_2=0$.  Upon reading out $g_1$ and $g_2$, a nucleus, a priori located in any of the four domains, can  unambiguously decide on a single one out of the four possibilities. This is equivalent to making 2 binary decisions and, as we will later show, to 2 bits of positional information.

Finally, the most subtle case is depicted in Fig.~\ref{ShapesANDCorr}C. Here, the mean profiles have exactly the same shapes as in Fig.~\ref{ShapesANDCorr}B. What is different, however, is the correlation structure of the fluctuations. In certain areas of the embryo the two genes are strongly positively correlated, while in the others they are strongly negatively correlated. If these areas are overlaid appropriately on top of the domains defined by the mean expression patterns, an additional increase in positional information is possible. In the admittedly contrived but pedagogical example of Fig.~\ref{ShapesANDCorr}C, the mean profiles and the correlations together define 8 distinguishable domains of expression, combinatorially encoded by 2 genes. Nuclei, having simultaneous access to the concentrations of the two genes, can compute which of the eight domains they reside in, although it might not be easy to implement such a computation in molecular hardware. Picking one of 8 choices corresponds to making 3 binary decisions, and thus to 3 bits of positional information. Note that in this case, each gene considered in isolation still carries 1 bit as before, so that the system of two genes carries more information than the sum of its parts---such a scheme is called synergistic encoding. 

In sum, we have shown that  the mean shapes of the profiles, as well as their variances \emph{and} correlations, can carry positional information. Extrapolating to 3 or more genes, we see that the number of pairwise correlations increases and in addition higher-order correlation terms start appearing. Formally, positional information could be encoded  in all of these features, but would become progressively harder to extract using plausible biological mechanisms. Nevertheless, a principled and assumption-free measure should combine all statistical structure into a single number, a scalar quantity measured in bits, that can ``count'' the number of distinguishable expression states (and thus positions), as illustrated in the examples above.

\subsubsection{Defining positional information}

Determining the number of  ``distinguishable states'' of gene expression is more complicated in real data sets than in our toy models: the mean profiles have complex shapes and their (co)variability depends on position. In the ideal case we can measure joint expression patterns of $N$ genes $\{g_i\},\;i=1,\dots, N$ (for example $N=4$ for four gap genes in \emph{Drosophila}) in a large set of embryos. In such a scenario the position dependence of the expression levels can be fully described with a conditional probability distribution $P(\{g_i\}|x)$. Concretely, for every position $x$ in the embryo we construct an $N$-dimensional histogram of expression levels across all recorded embryos, which (when normalized) yields the desired $P(\{g_i\}|x)$. This distribution contains all the information about how expression levels vary across embryos in a position-dependent fashion. For instance, 
\begin{eqnarray}
\bar{g}_i(x) &=& \int d^N\mathbf{g}\; g_i P(\{g_l\}|x), \label{mp} \\
\sigma_i^2(x) &=&\int d^N\mathbf{g}\; \left(g_i-\bar{g}_i(x)\right)^2P(\{g_l\}|x),\\
C_{ij}(x) &=&\int d^N\mathbf{g}\; \left(g_ig_j - \bar{g}_i(x)\bar{g}_j(x)\right)P(\{g_l\}|x), \label{covp}
\end{eqnarray}
are the mean profile of gene $g_i$, the variance across embryos of gene $g_i$, and the covariance between genes $g_i$ and $g_j$, respectively. In principle, the conditional distribution contains also all higher-order moments that we can extract by integrating over appropriate sets of variables. Realistically, we are often limited in our ability to collect enough samples to construct $P(\{g_i\}|x)$ by histogram counts, especially when considering several genes simultaneously; the number of samples needed grows exponentially with the number of  genes. However, estimating the mean profiles on the left-hand sides of Eqs.~(\ref{mp}-\ref{covp}) can often be   achieved from data directly. A reasonable first step (but one that has to be independently verified) is to assume that the joint distribution $P(\{g_i\}|x)$ of $N$ expression levels $\{g_i\}$ at a given position $x$ is Gaussian, which can be constructed using the measured mean values and covariances:

\begin{eqnarray}
\lefteqn{P(\{g_i\}|x)=(2\pi)^{-N/2}|C(x)|^{-1/2}\times}  \label{gaussian}\\
&& \times\exp\left[-\frac{1}{2}\sum_{i,j=1}^N(g_i-\bar{g}_i(x))[C^{-1}(x)]_{ij}(g_j-\bar{g}_j(x))\right] \nonumber
\end{eqnarray}
We emphasize that this Gaussian approximation is not required to theoretically define positional information, but that it will turn out to be practical when working with experimental data; in the cases of one or two genes it is often possible to proceed without making this approximation, which provides a convenient check for its validity.

While the conditional distribution, $P(\{g_i\}|x)$, captures the behavior of gene expression levels \emph{at a given $x$}, establishing how much  information, in total, the expression levels carry about position  requires us to know also how frequently each combination of gene expression levels, $\{g_i\}$, is used across all positions. Recall, for instance, that our arguments related to the information encoded in patterns of Fig.~\ref{ShapesANDCorr} rested on counting how often a pair of genes will be found in expression states 00, 01, 10, and 11 across all $x$. This global structure is  encoded in the total distribution of expression levels, which can be obtained by averaging the conditional distribution over all positions: 
\begin{equation}
P_g(\{g_i\})=\langle P(\{g_i\}|x)\rangle_x = \int_0^1 dx\; P(\{g_i\}|x), \label{Pg}
\end{equation}
where $\langle \cdot  \rangle_x$ denotes averaging over $x$.
Note that we can think of Eq.~(\ref{Pg}) as a special case of averaging with a position-dependent weight,
\begin{equation}
P_g(\{g_i\})=\int dx\; P_x(x) P(\{g_i\}|x),
\end{equation}
where $P_x(x)$ is chosen to be uniform. As we shall see, in the case of \emph{Drosophila} anterior-posterior (AP) patterning, $P_x(x)$ will be the distribution of possible nuclear locations along the AP axis, which indeed is very close to uniform. 

When formulated in the language of probabilities, the relationship between the position $x$ and the gene expression levels can be seen as a statistical dependency. If we knew this dependency were linear, we could measure it using, e.g., a linear correlation analysis between $x$ and $\{g_i\}$. Shannon has shown that there is an alternative measure of total statistical dependence (not just of its linear component), called the \emph{mutual information}, which is a functional of the probability distributions $P_x(x)$ and $P(\{g_i\}|x)$, and is defined by
\begin{eqnarray}
\lefteqn{ I(x\rightarrow \{g_i\})= } \nonumber \\
&&\int dx\;P_x(x) \int d^N \mathbf{g}\; P(\{g_i\}|x)\log_2\frac{P(\{g_i\}|x)}{P_g(\{g_i\})}. \label{mi1}
\end{eqnarray}
This positive quantity, measured in bits, tells us how much one can know about the gene expression pattern if one knows the position, $x$. It is not hard to convince oneself that the mutual information is symmetric, i.e., $I(\{g_i\}\rightarrow x)=I(x\rightarrow \{g_i\})=I(\{g_i\};x)$. This is very attractive: we do the experiments by sampling the distribution of expression levels given position, while the nuclei in a developing embryo implicitly solve the inverse problem---knowing a set of gene expression levels, they need to infer their position. A fundamental result of information theory states that both problems are quantified by the same symmetric quantity, the information  $I(\{g_i\};x)$. Furthermore, mutual information is not just one out of many possible ways of quantifying the total statistical dependency, but rather the unique way that satisfies a number of basic requirements, for example that information from independent sources is additive \cite{Shannon48,Cover91}.

The definition of mutual information in Eq.~(\ref{mi1}) can be rewritten as a difference of two entropies (which are always nonnegative):
\begin{equation}
I(\{g_i\};x)=S[P_g(\{g_i\})] - \langle S[P(\{g_i\}|x)] \rangle_x, \label{mi2}
\end{equation}
where $S[p(x)]$ is the standard entropy of the distribution $p(x)$ measured in bits (hence log base 2):
\begin{equation}
S[p(x)]=-\int dx\; p(x) \log_2p(x). \label{sdef}
\end{equation}

\begin{figure}
\centering
\includegraphics[width = 3.3in]{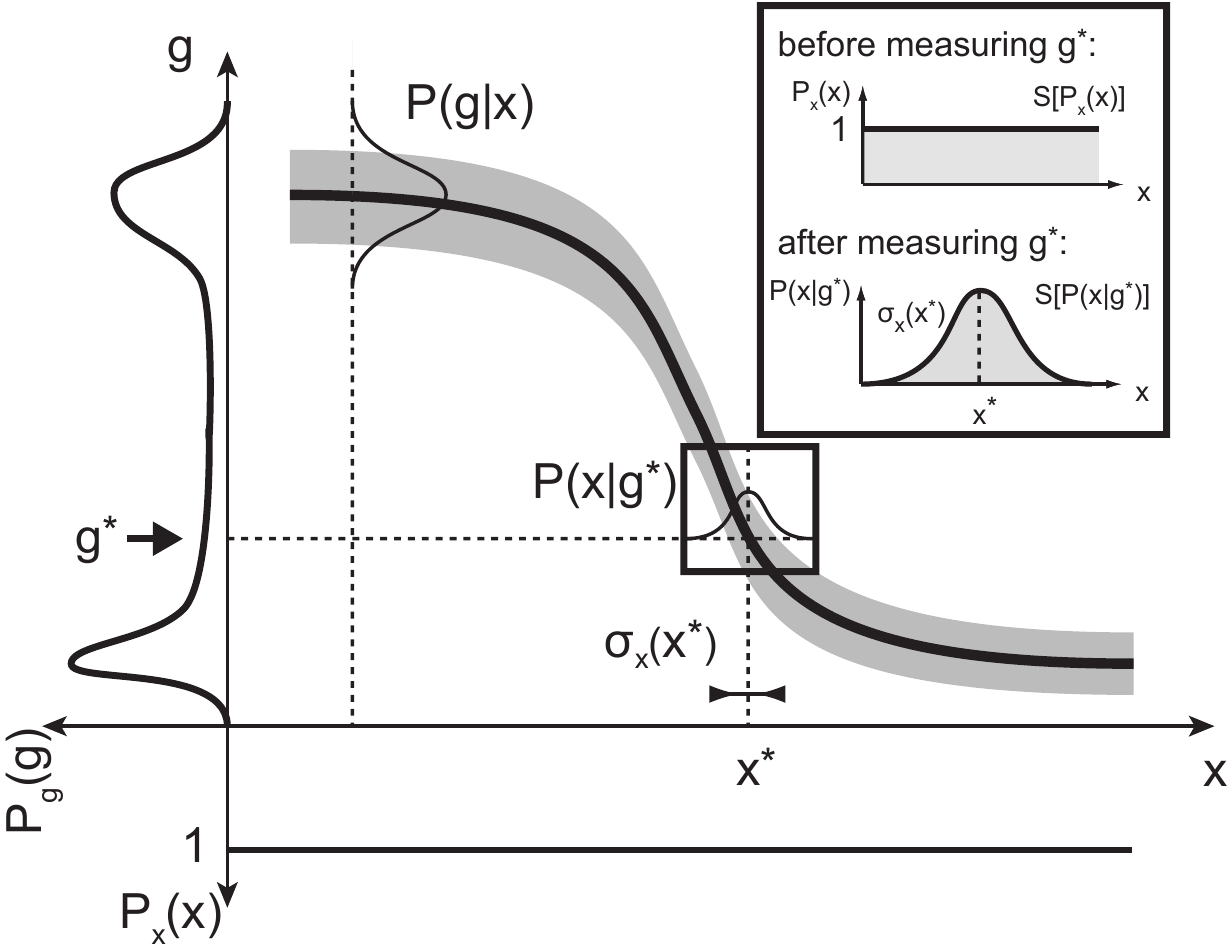}
\caption{\label{InfoDecoding} {\bf The mathematics of positional information and positional error for one gene.} A schematic  representation of the mean profile of gene $g$ (thick black line) and its variability (shaded envelope) across embryos. Nuclei are distributed uniformly along the AP axis, which is mathematically equivalent to saying that the prior distribution of nuclear positions  $P_x(x)$ is uniform (shown at the bottom).  For each position $x$, the gap gene expression levels $P(g|x)$ are, in this example, Gaussian. The total distribution of expression levels across the embryo, $P_g(g)$,  is determined by averaging $P(g|x)$ over all positions, and is shown on the left. The positional information $I(g;x)$ can be computed by averaging the difference of entropies of  $P_g(g)$ and $P(g|x)$ over all positions $x$, as in Eq.~(\ref{mi2}). {\bf Inset}: Decoding, or estimating the position of the nucleus, from a measured  expression level $g^*$. Prior to the ``measurement,'' all positions are equally likely. After observing the value $g^*$, the positions consistent with this measured value are drawn from $P(x|g^{*})$. The best estimate of the true position, $x^*$, is at the peak of this distribution, and the positional error, $\sigma_x(x)$, is the distribution's width. Due to the symmetry of mutual information, positional information $I(g;x)$ is also equal to the average difference between the entropy of the uniform distribution $P_x(x)$ and the entropy of $P(x|g)$.} 
\end{figure}

Equation~(\ref{mi2}) provides an alternative interpretation of the mutual information $I(\{g_i\};x)$ which is illustrated on the main panel of Fig.~\ref{InfoDecoding}. In the case of a single gene $g$, the ``total entropy'' $S[P_g(g)] $, represented on the left, measures the range of gene expression available across the whole embryo. This total entropy, or dynamic range, can be written as the sum of two contributions. One part is due to the systematic modulation of $g$ with position $x$, and this is the useful part  (the ``signal''), or the mutual information $I(\{g_i\};x)$. The other contribution is the variability in $g$ that remains even at constant position $x$;  this represents pure ``noise'' that carries no information about position, and is formally measured by the average entropy of the conditional distribution (the noise entropy), $\langle S[P(g|x)]\rangle_x$. Positional information carried by $g$ is thus the  difference between the total and noise entropies, as expressed in Eq.~(\ref{mi2}).

Mutual information is theoretically well founded, and is always non-negative, being 0 if and only if there is no statistical dependence of any kind between the position and the gene expression level. Conversely, if there are $I$ bits of mutual information between the position and the expression level, there are $\sim 2^{I(\{g_i\};x)}$ distinguishable gene expression patterns that can be generated by moving along the anterior-posterior (AP) axis, from the head at $x=0$ to the tail at $x=1$. This is precisely the property we require from any suitable measure of positional information. We therefore suggest that, mathematically, positional information should be defined as the mutual information between expression level and position, $I(\{g_i\};x)$. 

\subsubsection{Defining positional error}

Thus far we have discussed positional information in terms of the statistical dependency and the number of distinguishable levels of gene expression along the position coordinate. To present an alternative interpretation, we start by using the symmetry property  of the mutual information and  rewrite $I(\{g_i\};x)$ as
\begin{equation}
I(\{g_i\};x)=S[P_x(x)] - \langle S[P(x|\{g_i\})]\rangle_{P_g(\{g_i\})}, \label{mi3}
\end{equation}
i.e., the difference between the (uniform) distribution over all possible positions of a cell in the embryo, and the distribution of  positions consistent with a given expression level. Here, $P(x|\{g_i\})$ can be obtained using Bayes' rule from the known quantities:
\begin{equation}
P(x|\{g_i\}) = \frac{P(\{g_i\}|x)P_x(x)}{P_g(\{g_i\})}. \label{bayes}
\end{equation}
The total entropy of all positions, $S[P_x(x)]$ in Eq.~(\ref{mi3}), is independent of the particular regulatory system -- it simply measures the prior uncertainty about the location of the cells in the absence of knowing any gene expression level. If, however, the cell has access to the expression levels of a particular set of genes, this uncertainty is reduced, and it is hence possible to localize the cell much more precisely; the reduction in uncertainty is captured by the second term in Eq.~(\ref{mi3}). This  form of positional information emphasizes the \emph{decoding view,} that is, that cells can infer their positions by simultaneously reading out protein concentrations of various genes (Fig.~\ref{InfoDecoding}). 

Positional information is a single number: it is a global measure of the reproducibility in the patterning system. Is there a local quantity that would tell us, position by position, how well  cells can read out their gene expression levels and infer their location? Is  positional information ``distributed equally'' along the AP axis, or is it very non-uniform, such that cells in some regions of the embryo are much better at reproducibly assuming their roles?

The optimal estimator of the true location $x$ of a cell, once we (or the cell) measure the gene expression levels $\{g^{*}_i\}$, is the maximum a posteriori (MAP) estimate, $x^{*}(\{g^{*}_i\}) = \mathrm{argmax}\; P(x|\{g^{*}_i\})$. In cases like ours, where the prior distribution $P_x(x)$ is uniform, this equals the maximum likelihood (ML) estimate,
\begin{equation}
x^{*}(\{g^{*}_i\}) = \mathrm{argmax}\; P(\{g^{*}_i\}|x); \label{dec}
\end{equation}
thus, for each expression level readout, this ``decoding rule'' gives us the most likely position of the cell, $x^{*}$. The inset of Fig.~\ref{InfoDecoding} illustrates the decoding in the case of one gene.

How well can this (optimal) rule perform? The expected error of the estimated $x^{*}$ is given by $\sigma_x^2(x^{*})=\langle (x-x^{*})^2\rangle$, where brackets denote averaging over $P(x|\{g^{*}_i\})$. This error is a function of the gene expression levels; however, we can also evaluate it for every $x$, since we know the mean gene expression profiles, $\bar{g}_i(x)$, for every $x$. Thus, we define a new quantity, the \emph{positional error} $\sigma_x(x)$, which measures how well  cells at a true position $x$ are able to estimate their position based on the gene expression levels alone. This is the local measure of positional information that we were aiming for.

Independently of how cells actually read out the concentrations mechanistically, it can be shown that $\sigma_x(x)$  cannot be lower than the limit set by the Cramer-Rao bound \cite{Cover91}:
\begin{equation}
\sigma_x^2(x)\geq \frac{1}{\mathcal{I}(x)}, \label{perrorcr}
\end{equation}
where $\mathcal{I}(x)$ is the Fisher information given by
\begin{equation}
\mathcal{I}(x) = -\left\langle \frac{\partial^2 \log P(\{g_i\}|x)}{\partial^2x}\right\rangle_{P(\{g_i\}|x)}. \label{fisher}
\end{equation}
Despite its name, the Fisher information $\mathcal{I}$ is not an information-theoretic quantity, and unlike the mutual information $I$, the Fisher information depends on position. Is there a connection between the positional error, $\sigma_x(x)$, and the mutual information $I(\{g_i\};x)$? Below we sketch the derivation, following Ref.~\cite{Brunel98}, demonstrating the link for the case of one gene, $g$.

Let's assume that the Gaussian approximation of Eq.~(\ref{gaussian}) holds and that the distribution of the levels of a single gene at a given position is
\begin{equation}
P(g|x) = \frac{1}{\sqrt{2\pi \sigma_g^2(x)}}\exp\left\{-\frac{1}{2}\frac{(g-\bar{g}(x))^2}{\sigma_g^2(x)}\right\}.
\end{equation}

We can use the Gaussian distribution to compute the Fisher information in Eq~(\ref{fisher}). We find that
\begin{equation}
\mathcal{I} = \frac{\bar{g}'^2(x)}{\sigma_g^{2}(x)} + 2\frac{\sigma_g'^2(x)}{\sigma_g^2(x)},
\end{equation}
where $(\cdot)'$ denotes a derivative with respect to position, $x$. Information about position is thus carried by the change in mean profile with position, as well as the change in the variability itself with position. If the noise is small, $\sigma_g\ll \bar{g}$, we can retain only the first term to obtain a bound on positional error:
\begin{equation}
\sigma_x^2(x)\geq \frac{1}{\mathcal{I}(x)} \approx \left(\frac{d\bar{g}}{dx}\right)^{-2} \sigma_g^2(x). \label{sigmax1}
\end{equation}
This result is intuitively straightforward: it is simply the transformation of the variability in gene expression, $\sigma_g^2(x)$, into an effective variance in the position estimate, $\sigma_x^2(x)$, and the two are related by slope of the input/output relation, $\bar{g}(x)$. 

A crucial next step is to think of $x$ as determining gene expressions $g_i$ probabilistically, and the $x^*$ as being a function of these gene expression levels---that is, when computing $x^*$ neither we nor the nuclei have access to the true position. This forms a dependency chain, $x\rightarrow\{g_i\}\rightarrow x^*$. Since each of these steps is probabilistic,  it can only lose information, such that by information processing inequality \cite{Cover91} we must have $I(\{g_i\};x) \geq I(x^*;x)$. The mutual information between the true location and its estimate is given by  
\begin{equation}
I(\hat{x};x) = S[P_x(x^*)] - \langle S[P(x^*|x)]\rangle_{P_x(x)}.
\end{equation}
Under  weak assumptions, the first term in our case is approximately the entropy of a uniform distribution. While we don't know the full distribution $P(x^*|x)$ and thus cannot compute its entropy directly, we know its variance, which is just the square of the positional error, $\sigma_x^2(x)$. Regardless of what the full distribution is, its entropy must be less or equal to the entropy of the Gaussian distribution of the same variance, which is $S[P(x^*|x)]=\log_2\sqrt{2\pi e \sigma_x^2(x)}$ bits. Putting everything together, we find that:
\begin{equation}
I(\{g_i\};x) \geq I(x^*;x) \geq \Big\langle \log_2 \frac{P_x(x)}{\sqrt{2\pi e \sigma_x^2(x)}}  \Big\rangle_x. \label{ineq}
\end{equation}
Therefore, positional information $I(\{g_i\};x)$ puts an upper bound to the average ability of the cells to infer their locations, that is, to the smallness of the positional error $\sigma_x(x)$. In a straightforward generalization of a single gene case, the Fisher information for a multi-variate Gaussian distribution, written for compactness in matrix notation, yields:
\begin{equation}
\mathcal{I} = \bar{\mathbf{g}}'^T \mathbf{C}^{-1} \bar{\mathbf{g}}' + \frac{1}{2}\mathrm{Tr} \left[ \mathbf{C}^{-1} \mathbf{C}' \mathbf{C}^{-1} \mathbf{C}' \right]. \label{fullFI}
\end{equation}
Under the same assumption of small noise that we made for the case of a single gene, we retain only the first term, so that the expression for positional error, written out explicitly in the component notation, reads:
\begin{equation}
\sigma_x^2(x) \geq \frac{1}{\mathcal{I}(x)}\approx \left(\sum_{i,j=1}^N \frac{d\bar{g}_i}{dx} [C(x)^{-1}]_{ij} \frac{d\bar{g}_j}{dx}\right)^{-1}, \label{poserr}
\end{equation}
where $C_{ij}$ is the covariance matrix of the profiles, as defined in Eq.~(\ref{covp}). This extends the fundamental connection, Eq.~(\ref{ineq}), between the positional information and positional error, to the case of multiple genes.  Importantly, all quantities---the mean profiles and their covariance---in Eq.~(\ref{poserr}) can be obtained from experimental data, so $\sigma_x(x)$ is a quantity that can be estimated directly.

\subsubsection{Interpretation in a spatially discrete (cellular) system}

In the setup presented above, gene expression levels carry information about a continuous position variable, $x$. But in a real biological system, there exists a minimal spatial scale---the scale of individual nuclei (or cells)---below which the concept of gene expression \emph{at a position} is no longer well defined. This is particularly the case in systems that interpret molecular concentrations to make decisions that determine cell fates: such interpretations have no meaning at the spatial scale of a molecule, but only at cellular scales. How should positional information be interpreted in such a context?

\begin{figure}
\centering
\includegraphics[width = 3.3in]{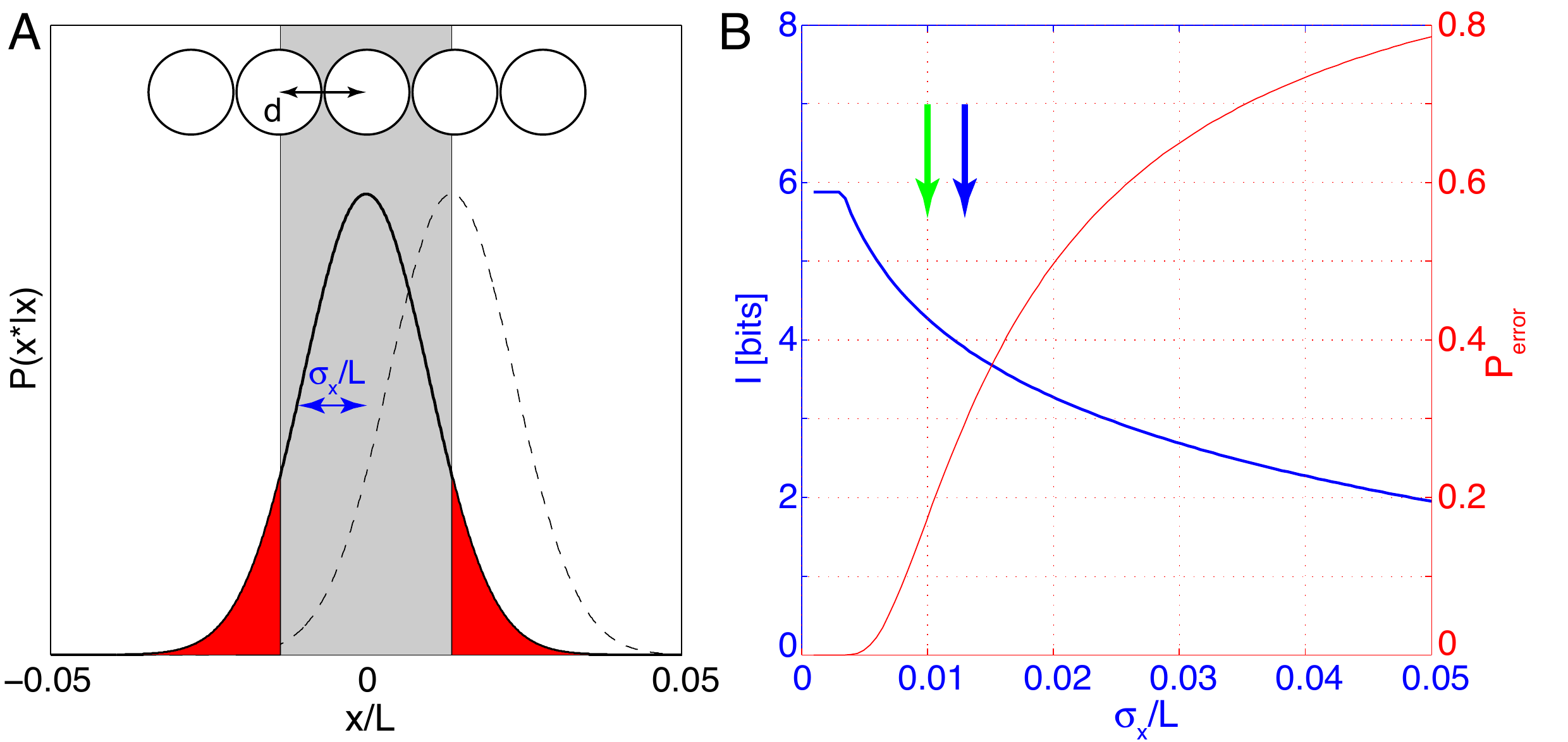}
\caption{{\bf Positional information, error, and perfect identifiability.} {\bf A)} A row of nuclei, separated by distance $d$, decode their position from gap gene expression levels (top). The probability $P(x^*|x)$ for the estimated position $x^*$ is Gaussian (solid black line for the central nucleus and dashed line for the right next nucleus); its width is the positional error $\sigma_x$. The probability $P_{\rm error}$ that the central nucleus is assigned to the wrong lattice position is equal to the integral of the tails $|x^*|>d$ of the distribution $P(x^*|x)$ (red area). {\bf B)} Positional information (blue) and probability of false  assignment (red) as a function of positional error $\sigma_x$. Blue arrow indicates the value of positional error $\sigma_x=d$ used for the toy example in A; specifically we used $N_n=59$ nuclei uniformly tightly-packing the central 80\% of the AP axis ($L=0.8$ and thus $d=L/N_n$). In comparison, the 1\% positional error inferred from data in Ref \cite{PNASPI} corresponds to the green arrow. The information is computed using Eq.~(\ref{infox}) and made to saturate at $I_{\rm max}=\log_2(N_n)$ bits (perfect identifiability). All parameters are chosen to roughly match the results of the \emph{Drosophila} gap gene analysis.} 
\label{f4}
\end{figure}

When noise is small enough and the distribution of positions consistent with observed gene expression levels, $P(x|\{g_i\})$ of Eq.~(\ref{bayes}), is nearly Gaussian, Eqs.~(\ref{ineq} and \ref{poserr}) become tight bounds. In this case we can apply Eq.~(\ref{mi3}) directly. In developmental systems, cells or nuclei are often distributed in space such that the inter-cellular (or inter-nuclear) spacings are, to a good approximation, equal: at least in systems we are studying, there are no significant local rarefaction or overabundances of cells or nuclei. Mathematically, this amounts to assuming that  $P_x(x)=1/L$ (where $L$ is the linear spatial extent over which the cells or nuclei are distributed, with  $L=1$ in our convention). This assumption of uniformity is not crucial for any calculation in this paper and can be easily relaxed, but it makes the equations somewhat simpler to display and interpret. Using an uniform distribution for $P_x(x)$, the information is: 
\begin{equation}
I(\{g_i\};x)\approx \big\langle\log_2(L/\sqrt{2\pi e \sigma_x^2(x)})\big\rangle_x. 
\label{infox}
\end{equation}
For any real dataset one needs to verify that  the approximations leading to this result are warranted (see below). Assuming that, Eq.~(\ref{infox}) further illustrates the connection between positional error and positional information. Consider a one-dimensional row of nuclei spaced by inter-nuclear distance $d$ along the AP axis, as illustrated in Fig.~\ref{f4}.  In order to determine its position along the AP axis, each nucleus has to generate an estimate $x^*$ of its true position $x$ by reading out gap gene expression levels. In the simplest case, the errors of these estimates are independent, normally distributed, with a mean of zero and a variance $\sigma_x^2$. Given these parameters, there is some probability $P_{\rm error}$ that the positional estimate deviates by more than the lattice spacing $d$, in which case the nucleus would be assigned to the wrong position, and perfectly unique, nucleus-by-nucleus identifiability would be impossible. Figure~\ref{f4} shows how the positional information of Eq.~(\ref{infox}), and the probability of fate misassignment, $P_{\rm error}$, vary as functions of $\sigma_x$. Importantly, even when $\sigma_x<d$, that is, the positional error is smaller than the inter-nuclear spacing (as in Fig.~\ref{f4}A), there is still some probability of nuclear misidentification and therefore the positional information has not yet saturated. Only when the positional error is sufficiently small that the probability weight in the tails of the Gaussian distribution is negligible (at $\sigma_x\ll d$), can each of the $N_{n}$ nuclei be perfectly identified and the information saturates at $\log_2(N_n)$ bits.

In sum, we have shown that a rigorous mathematical framework of positional information can quantify the reproducibility of gene expression profiles in a global manner. By framing the cells' problem of finding their location in the embryo in terms of an estimation problem, we have shown that the same mathematical framework of positional information places precise constraints on how well the cells can infer their positions by reading out a set of genes. These constraints are universal: regardless of how complex the mechanistic details of the cells' readout of the gene concentrations levels are, the expression level variability prevents the cells from decreasing the positional error below $\sigma_x$. The concept of  positional error  easily generalizes to the case of multiple genes, and is diagnostic about how positional information is distributed along the AP axis.

\subsection{Technical challenges}

In this section, we survey some technical details related to the application of the formalism of positional information and positional error to real data sets. Estimating information theoretic quantities from finite data is, generally, a very difficult problem that has received considerable theoretical attention (see, e.g., \cite{paninski,Slonim05}). The way we approach this problem productively is to combine general theory and algorithms for information estimation with problem-specific approximations. We use the four major gap genes in early \emph{Drosophila} embryos as a test case. First, we briefly recapitulate our experimental and data processing methods (for details, see Ref~\cite{MSB}). Next, we present the statistical techniques necessary to consistently merge data from separate pairwise gap gene immunostaining experiments into a single dataset. We estimate mutual information directly from data for one gene using different profile normalization and alignment methods. Finally, we introduce the Gaussian noise approximation and an adaptive Monte Carlo integration scheme in order to extract the information carried jointly by pairs of genes and the full set of four genes.

\subsubsection{Extracting expression level profiles from imaging data}

\emph{Drosophila} embryos were fixed and simultaneously immunostained for the four major gap gene [\emph{hunchback} (\emph{hb}), \emph{kr\"uppel} (\emph{kr}), \emph{knirps} (\emph{kn}) and \emph{giant} (\emph{gt})] proteins, using fluorescent antibodies with minimal spectral overlap, as described in Ref~\cite{MSB}. In this paper, we analyze a total of four datasets (A, B, C and D). Datasets A--C have been processed simultaneously, but imaged in different sessions. Dataset D has been processed and imaged independently from the other datasets. Hence these datasets are well suited to assess the dependence of our measurements and calculations on the experimental processing. Dataset A has been used previously in Refs~\cite{MSB,PNASPI,Dima}. 

Fluorescence intensities were measured using automated laser scanning confocal microscopy, processing hundreds of embryos in one imaging session. Cross-sectional images of multi-color labeled embryos were taken with an imaging focus at the midsagittal plane, the center plane of the embryo with the largest circumference. Intensity profiles of individual embryos were extracted along the outer edge of the embryos using custom software routines (MATLAB, MathWorks, Natick, MA) as described \cite{Houchmandzadeh02, MSB}. The results in the following sections are presented using exclusively dorsal intensity profiles, and we report their projection onto the anteroposterior (AP) axis in units normalized by the total length $L$ of the embryo, yielding a fractional coordinate between 0 (anterior) and 1 (posterior). The dorsal edge was chosen for its smaller curvature, thus limiting geometric distortion of the profiles when projecting the intensity onto the AP axis. The AP axis was uniformly divided into 1000 bins, and the average profile intensity in each bin is reported. This procedure results in a raw data matrix that lists, for each embryo and for each of the four gap genes, one intensity value for each of the 1000 equally spaced spatial positions along the AP axis.

For any successful information-theoretic analysis, the dominant source of observed variability in the dataset must be due to the biological system and not due to the measurement process, requiring tight control over the experimental setup. Gap gene expression levels critically depend on the developmental stage and on the imaging orientation of the embryo. Since multiple embryos need to be pooled together to assess embryo-to-embryo fluctuations, precise control of these two factors is necessary in order to maximally reduce systematic variability that can be attributed to measurement noise. Therefore we applied strict selection criteria on developmental timing and orientation angle (see Ref. \cite{MSB}). In this paper we restrict our analysis to a time window of $\sim\!\!10\min$, 38--$48\min$ into nuclear cycle 14. During this time interval the mean gap gene expression levels peak and overall temporal changes are minimal. A carful analysis of residual variabilities due to measurement error (i.e., age determination, orientation, imaging, antibody non-specificity, spectral cross-talk, and focal plane determination) reveals that the estimated fraction of observed variance in the gap gene profiles due to systematic and experimental error is below 20\% of the total variance in the pool of profiles, i.e., more than 80\% of the variance is due to the true biological variability in the gap gene system \cite{MSB}. The fact that the overwhelming fraction of the total variability in our data sets is due to natural fluctuations is a prerequisite for the proposed information-theoretic analysis to yield biological insights.

Before the profiles can be compared or aggregated across embryos, they may need to be normalized. Careful experimental design and imaging can make normalization steps essentially superfluous \cite{MSB}, but such control may not always be possible. In general, different antibodies could have different overall specificities and spectral efficiency, and  there could be small embryo-to-embryo variations in the overall fluorescence background and staining penetration, even if all the embryos are prepared and imaged concurrently. To deal with these artifacts, we introduce three possible types of normalizations, which we call {\tt Y} alignment, {\tt X} alignment, and {\tt T} alignment. In {\tt Y} alignment, the recorded intensity of immunostaining for each profile of a given gap gene, $G^{(\mu)}(x)$ recorded from embryo $\mu$ ($\mu=1,\dots,\mathcal{N}$), is assumed to be linearly related to the (unknown) true concentration profile $g^{(\mu)}(x)$ by an additive constant $\alpha_\mu$ and  an overall scale factor $\beta_\mu$ (to account for the background and staining efficiency variations from embryo to embryo, respectively), so that $G^{(\mu)}(x) = \alpha_\mu + \beta_\mu g^{(\mu)}(x)$.  We would like to minimize the total deviation $\chi^2$ of the concentration profiles from the mean across all embryos. The objective function is 
\begin{equation}
\chi_{\mathrm Y}^2(\{\alpha_\mu,\beta_\mu\}) = \sum_{\mu=1}^{\mathcal{N}} \int_{0}^1dx\; \big( G^{(\mu)}(x)  - \left[\alpha_\mu + \beta_\mu  \bar{g}(x)\right]  \big)^2,
\end{equation}
where $ \bar{g}(x)=\mathcal{N}^{-1}\sum_\mu g^{(\mu)}(x)$ denotes an average concentration profile across all $\mathcal{N}$ embryos. The parameters $\{\alpha_\mu,\beta_\mu\}$ are chosen to minimize the $\chi_Y^2$ and thus maximize alignment, and since the cost function is quadratic, this optimization has a closed form solution \cite{Gregor07a,MSB}.

Additionally, one can perform the {\tt X} alignment, where \emph{all} gap gene profiles from a given embryo can be translated along the AP axis by the same amount; this introduces one more parameter $\gamma_i$ per embryo (\emph{not} per expression profile), which can be again determined using $\chi^2$ minimization, similar to the above \footnote{In practice, one first carries out the tractable {\tt Y} alignment, followed by a joint minimization of $\chi^2$ for $\alpha,\beta,\gamma$, which needs to be carried out numerically.}. There are two candidate sources of variability that can be compensated for by {\tt X} alignments: (i) our error in the exact determination of the AP axis, i.e., in the exact end-points of the embryo, due to image processing; (ii) real biological variability that would result in all the nuclei within the embryo being rigidly displaced by a small amount along the long axis of the embryo relative to the egg boundary. As we will see in Discussion, the second option has an interesting biological interpretation in terms of correlated positional errors that nuclei might make while reading out positional information.

Lastly, the {\tt T} alignment attempts to compensate for the fraction of observed variability that is due to the systematic change in gene profiles with embryo age even within the chosen 10 minute time bracket. We can use our knowledge of how the mean profiles evolve with time and de-trend the entire dataset in a given time window by this evolution of the mean profile. We follow the procedure outlined in Ref~\cite{MSB} to carry out this alignment. 

In sum, each of the three alignments, {\tt X}, {\tt Y} and {\tt T}, subtracts  from the total variance the components which are likely to have an experimental origin and  don't represent properly either the intrinsic noise within an embryo or natural variability between the embryos; since the total variance will be lower after alignment, successive alignment procedures should lead to increases in positional information. Strictly speaking, the lower bound on positional information would be obtained by estimating it using raw data, without any alignment, thus ascribing \emph{all} variability in the recorded profiles to the true biological variability in the system, but unless the control over experimental variability is excellent, this lower bound might be far below the true value. For instance, if embryos cannot be stained and imaged in a single session, it would be very hard to guarantee that there are no embryo-to-embryo variabilities in the antibody staining and imaging background. For that reason, we view the {\tt Y} alignment as the minimal procedure that should be performed unless staining and imaging variability is shown to be negligible in dedicated control experiments. The choice of performing  alignments beyond {\tt Y} depends on system-specific knowledge about the plausible sources of experimental vs biological variability. For these reasons, most of the results in the paper are based on the minimal alignment procedure ({\tt Y}), thus yielding conservative information estimates, but we also explore how these estimates would increase for single genes and the quadruplet in case of the other alignments.

Once the desired alignment procedure has been carried out, we can define the mean expression across the embryos, $\bar{g}_i(x)$ for every gap gene $i=1,\dots,4$, and choose the units for the profiles such that the minimum value of each mean profile across the AP axis is 0, and the maximal value is 1. This is the final, aligned and normalized, set of profiles on which we carry out all subsequent analyses, and from which we compute the covariance matrix $C_{ij}(x)$ of Eq~(\ref{covp}).

\begin{figure}
\centering
\includegraphics[width = 3.3in]{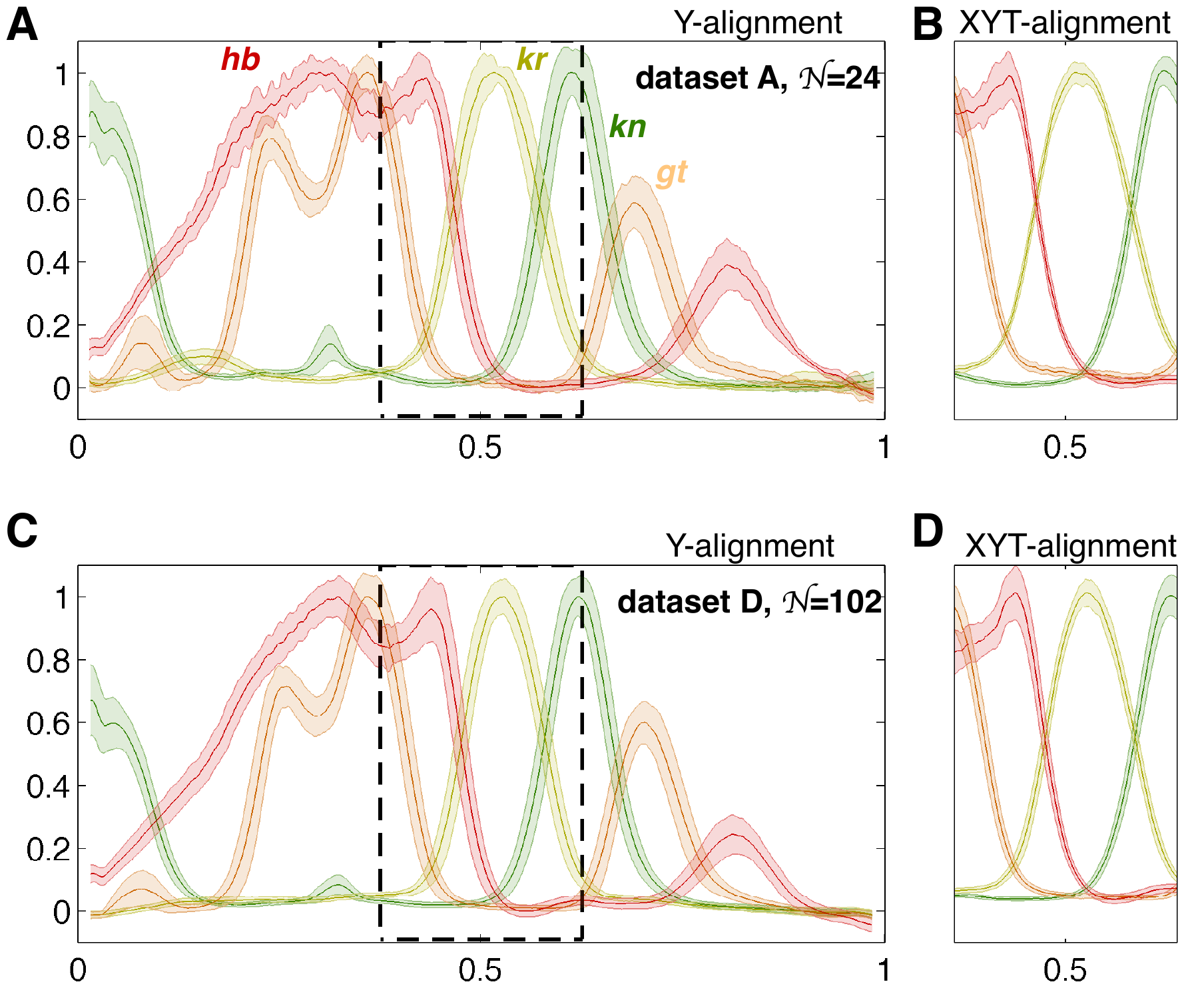}
\caption{\label{f5} {\bf Simultaneous measurements of gap gene expression levels and profile alignment.} {\bf A)} The mean profiles (thick lines) of gap genes Hunchback, Giant, Knirps, Kr\"uppel (color coded as indicated) and the profile variability  (shaded region = $\pm 1$ std envelope) across 24 embryos in data set A, normalized using the {\tt Y} alignment procedure. {\bf B)} The same profiles have been aligned using the full ({\tt XYT}) procedure. Shown is the region that corresponds to the dashed rectangle in A to illustrate the substantially reduced variability across the profiles. {\bf C, D)} Plots analogous to A, B showing the profiles and their variability for Dataset D.  }
\end{figure}

Applying these selection criteria to the four mentioned datasets yields embryo counts of $\mathcal{N}=24$ (A), $\mathcal{N}=32$ (B), $\mathcal{N}=31$ (C), and $\mathcal{N}=102$ (D). Figure~\ref{f5} shows the mean profiles and their variability for two of the datasets using either the minimal ({\tt Y}) or full ({\tt XYT}) alignment.

\subsubsection{Estimating information with  limited amounts of data}

Measuring positional information from a finite number $\mathcal{N}$ of embryos is challenging due to estimation biases.  Good estimators are thus often more complicated than the naive approach, which consists of estimating the relevant distributions by counting and  using the formula, e.g. Eq~(\ref{mi1}), for mutual information directly. Nevertheless, the naive approach can be used as a basis for an unbiased information estimator following the so-called \emph{direct method} \cite{Strong98,Slonim05}.

\begin{figure}
\centering
\includegraphics[width = 3.3in]{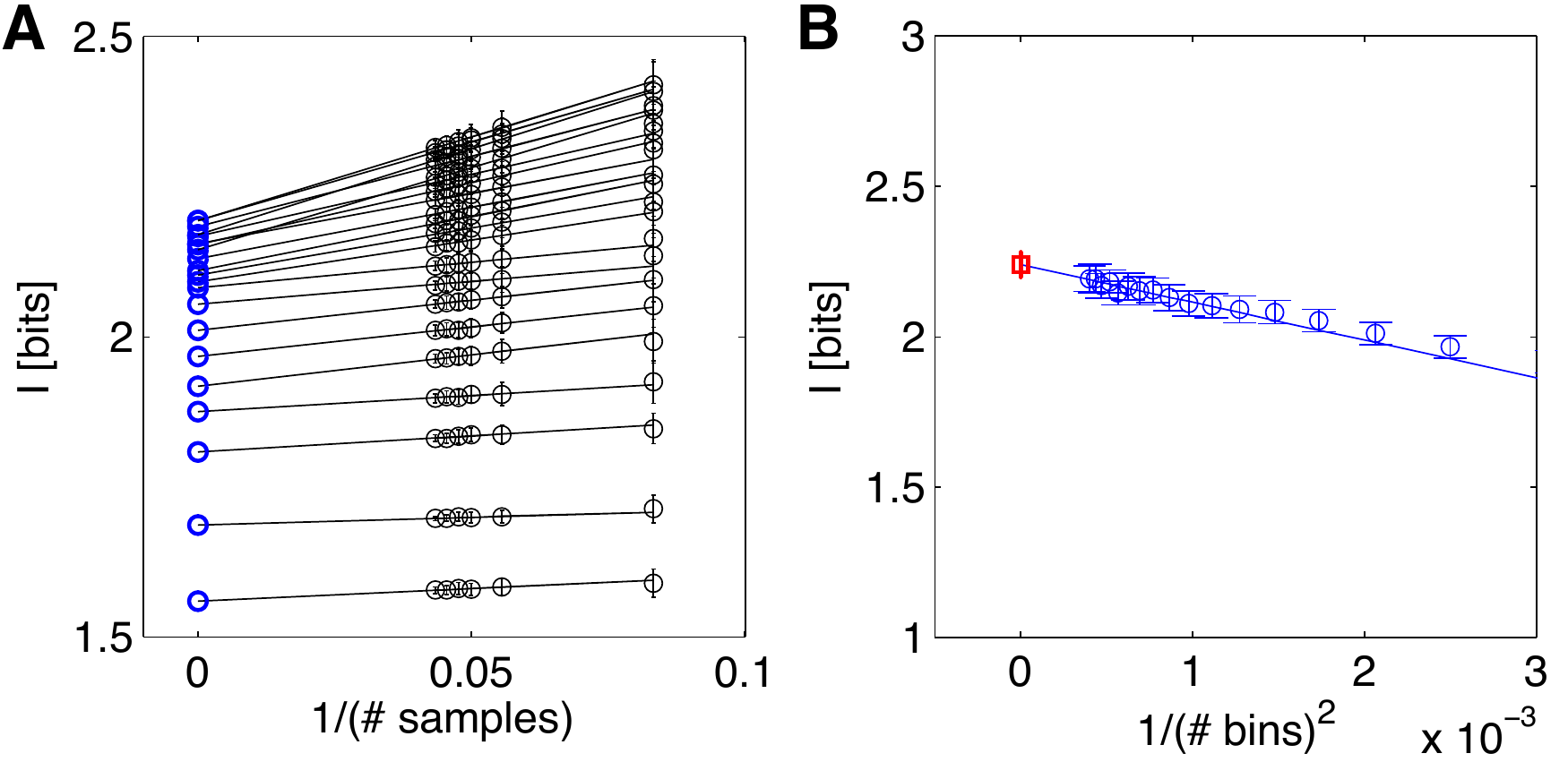}
\caption{\label{f6} {\bf Direct estimation for the positional information carried by Hunchback in the 10-90\% egg length segment from Dataset A  using {\tt Y} alignment.} {\bf A)} The first extrapolation of the direct method consists of selecting different subsets of the embryos, and performing a series of naive estimates using different bin sizes $\Delta$. Each choice of bin size corresponds to one extrapolation in the plot (starting with 10 bins for the bottom line and increasing to 50 bins for the top line in increments of 2). Black points are averages of naive estimates over 100 random choices of $m$ embryos (error bars = std), plotted against $1/m$ on the x-axis. Lines and blue circles represent extrapolations to the infinite data limit, $m\rightarrow\infty$. {\bf B)} The extrapolations to infinite data limit (blue points from A) are plotted as a function of the bin size, and the second regression is performed (blue line) to find the extrapolation of $\Delta\rightarrow 0$. The final estimate, represented by the red square, is $I^{\rm DIR}(hb;x)=2.26\pm 0.04$ bits; the error bar is the statistical uncertainty in the extrapolated value due to limited sample size.}
\end{figure}

The easiest way to obtain a naive estimate for $P(\{g_i\},x)$ is to convert the range of continuous values for $g_i$ and $x$ into discrete bins of size $\Delta^N \times \Delta$. On this discrete domain, we can estimate the distribution  $\tilde{P}_{\Delta,M} (\{g_i\},x)$ empirically by histogramming, treating our data matrix of  (4 gap genes) $\times$ ($\mathcal{N}$ embryos) $\times$ (1000 spatial bins) as containing $M=1000\times \mathcal{N}$ samples from the joint distribution of interest.    A \emph{naive estimate} of the positional information, $I_{\Delta,M}^{\textrm{DIR}}(\{g_i\};x)$ is:
\begin{eqnarray}
&& I_{\Delta,M}^{\textrm{DIR}}(\{g_i\};x) =   \\
 &&\!\! \sum_{\{g_i\},x} \, \tilde{P}_{\Delta,M}(\{g_i\},x)\log_2 \frac{ \tilde{P}_{\Delta,M}(\{g_i\},x) }{ \tilde{P_x}_{\Delta,M}(x)  \tilde{P_g}_{\Delta,M}(\{g_i\})} \nonumber
\end{eqnarray}
where the subscripts indicate the explicit dependence on sample size $M$ and bin size $\Delta$.
It is known that naive estimators suffer from estimation biases that scale as $1/M$ and $\Delta^{N+1}$. Following Refs.~\cite{Strong98,Slonim05}, we can obtain a \emph{direct estimate} of the mutual information by first computing a series of naive estimates for a fixed value of $\Delta$ and for fractions of the whole data set. Concretely, we pick fractions $m=[0.95\; 0.9\; 0.85\; 0.8\; 0.75\; 0.5]\times \mathcal{N}$ of the total number of embryos, $\mathcal{N}$. At each fraction, we randomly pick $m$ embryos 100 times and compute $\langle I_{\Delta, m}^{\textrm{DIR}}(\{g_i\};x)\rangle$ (where averages are taken across 100 random embryo subsets). This gives us a series of data points that can  be extrapolated to an infinite data limit by linearly regressing $\langle I_{\Delta, m}^{\textrm{DIR}}(\{g_i\};x)\rangle$ vs $1/m$ (Fig.~\ref{f6}A). The intercept of this linear model yields $I_{\Delta, m\rightarrow\infty}^{\textrm{DIR}}(\{g_i\};x)$, and we can repeat this procedure for a set of ever smaller bin sizes $\Delta$. To extrapolate the result to  very small bin sizes, $\Delta\rightarrow 0$, we use the previously computed $I_{\Delta, \infty}^{\textrm{DIR}};(\{g_i\};x)$ for various choices of decreasing $\Delta$, and extrapolate to $\Delta\rightarrow 0$ as shown in Fig.~\ref{f6}B. At the end of this procedure we obtain the final estimate $I_{\Delta\rightarrow 0, n\rightarrow\infty}^{\textrm{DIR}}(\{g_i\};x)$, called \emph{direct estimate (DIR)} of positional information (red square in Fig.~\ref{f6}B). While no prior knowledge about the shape of the distribution $P(\{g_i\},x)$ is assumed by the direct estimation method, a potential disadvantage is the amount of data required, which grows exponentially in the number of gap genes. In practice, our current data sets suffice for the direct estimation of positional information carried simultaneously by one or at most two gap genes.

To extend this method tractably to  more than two genes, one needs to resort to approximations for $P(\{g_i\}|x)$, the simplest of which is the so-called Gaussian approximation, shown in Eq.~(\ref{gaussian}). In this case, we can write down the entropy of $P(\{g_i\}|x)$ analytically. For a single gene we get
\begin{equation}
S[\tilde{P}_{\Delta,M}(g|x)]=\frac{1}{2}\log_2\left(2\pi e\sigma_g^2(x)\right) + \log\Delta, \label{fga}
\end{equation}
while the straightforward generalization to the case of $N$ genes is given by
\begin{equation}
S[\tilde{P}_{\Delta,M}(\{g_i\}|x)] = \frac{1}{2}\log_2\left((2\pi e)^N |\mathbf{C}(x)|\right) + N\log\Delta, \label{fgaN}
\end{equation}
where $|\mathbf{C}(x)|$ is the determinant of the covariance matrix $C_{ij}(x)$. 
From Eq.~(\ref{mi2}) we know that  $I(g;x)=S[P(g)]-\langle S[P(g|x)]\rangle_x$. Here the second term (``noise entropy'') is therefore easily computable from Eq~(\ref{fga}), for a discretized version of the distribution, $\tilde{P}$, using the (co-)variance estimate of gene expression levels alone. The first term (``total entropy'') can be estimated as above by the direct method, i.e. by histogramming $\tilde{P}_{\Delta,M}(g)$ for various sample sizes $M$ and bin sizes $\Delta$, and extrapolating $M\rightarrow\infty$ and $\Delta\rightarrow 0$. This combined procedure, where we evaluate one term in the Gaussian approximation and the other one directly, has two important properties. First, the total entropy is usually much better sampled than the noise entropy, because it is based on the values of $g$ pooled together over every value of $x$; it can therefore be estimated in a direct (assuption-free) way even when the noise entropy cannot be. Second, by making the Gaussian approximation for the second term, we are always \emph{overestimating} the noise entropy and thus \emph{underestimating} the total positional information, because the Gaussian distribution is the maximum entropy distribution with a given mean and variance (other, non-Gaussian distributions with the same mean and variance can only have smaller entropies). Therefore, in a scenario where the first term in Eq.~(\ref{fga}) is estimated directly, while the second term is computed analytically from the Gaussian ansatz, we always obtain a lower bound on the true positional information. We call this bound (which is tight if the conditional distributions really are Gaussian) the \emph{first Gaussian approximation} (FGA).

\begin{figure*}
\centering
\includegraphics[width = 6.0in]{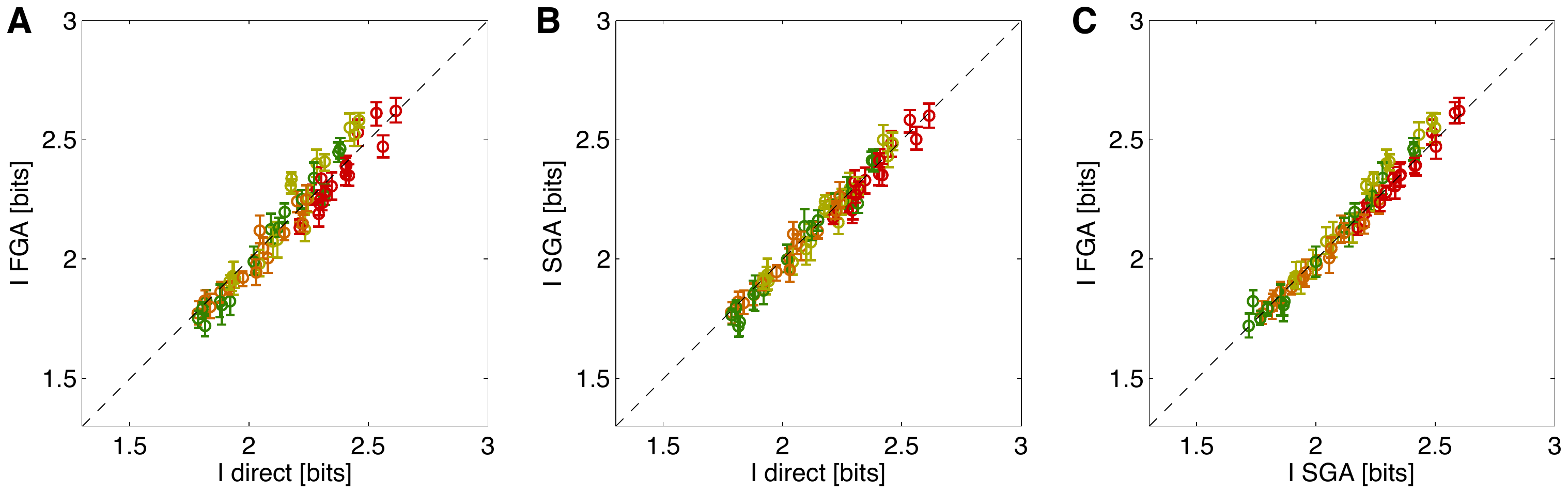}
\caption{\label{f7}{\bf Comparing estimation methods for positional information carried by single gap genes in the 10-90\% egg length segment.} Shown are the estimates for 4 gap genes (color coded as in Fig~\ref{f5}), using 4 datasets (Datasets A, B, C, D) and 4 different alignment methods ({\tt Y}, {\tt YT}, {\tt XY}, {\tt XYT}), for 64 total points per plot. Dashed line shows equality. Different alignment methods result in a spread of information values for the same gap gene, as shown explicitly Fig~\ref{f8}.  }
\end{figure*}

For  three or more gap genes the amount of data can be insufficient to reliably apply either the direct estimate or FGA, and one needs to resort to yet another approximation, called  the \emph{second Gaussian approximation} (SGA). As in the FGA, for the second Gaussian approximation we also assume that $P(\{g_i\}|x)\approx\mathcal{G}(\{g_i\};\bar{g}_i(x), C_{ij}(x))$ is Gaussian, but we make another assumption in that $P_g(\{g_i\})$, obtained by integrating over these Gaussian conditional distributions, is a good approximation to the true $P_g(\{g_i\})$. The total distribution we use for the estimation is therefore a Gaussian mixture:
\begin{equation}
P_g(\{g_i\})=\int_0^1dx\; P(\{g_i\}|x). \label{int2nd}
\end{equation}
For each position, the noise entropy in Eq.~(\ref{fgaN}) is proportional to the logarithm of the determinant of the covariance matrix, which scales as $1/M$ if estimated from a limited number $M$ of samples. We therefore estimate the information $I_{M}^{\textrm{SGA}}(\{g_i\};x)$ for fractions of the whole data set and then extrapolate for $M\rightarrow\infty$. 

Figure~\ref{f7} compares the three methods for computing the positional information carried by single gap genes in the 10-90\% egg length segment. Across all four gap genes and all four datasets the methods agree within the estimation error bars. This provides implicit evidence that, at least on the level of single genes, the Gaussian approximation holds sufficiently well for our estimation methods.

\begin{figure}[b]
\centering
\includegraphics[width = 3.2in]{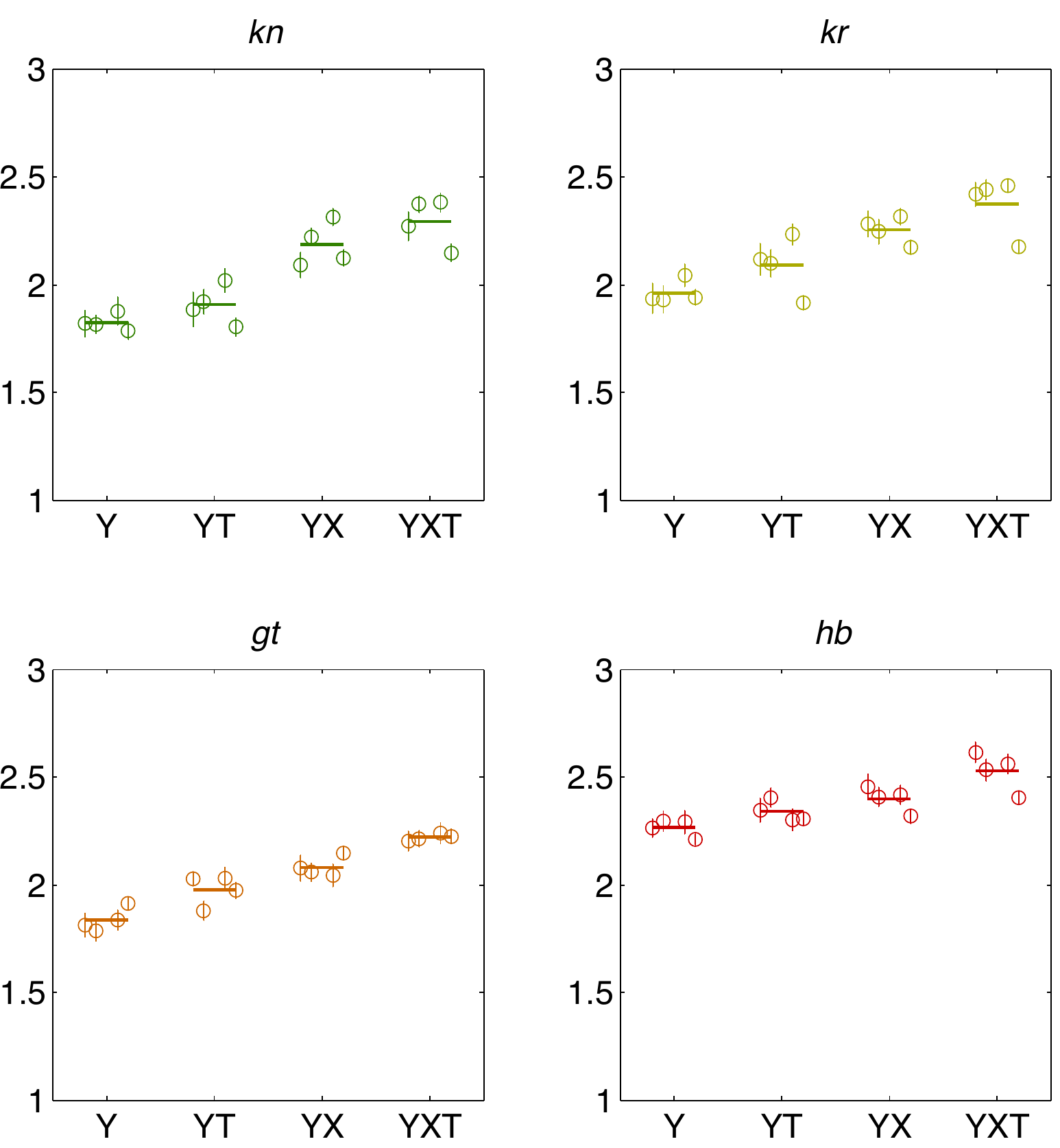}
\caption{\label{f8}{\bf Consistency across datasets and a comparison of alignment methods for positional information carried by single gap genes.} Direct estimates with estimation error bars are shown for 4 gap genes (color coded as in Fig~\ref{f5}) in separate panels. On each panel, different alignment methods ({\tt Y}, {\tt YT}, {\tt XY}, {\tt XYT}) are arranged on the horizontal axis, and for every alignment method we report the estimation results using Datasets A, B, C, D (successive plot symbols) and their average (thick horizontal line).  }
\end{figure}

Figure~\ref{f8} compares the estimation results across datasets and the alignment methods. With the exception of Dataset D data under {\tt T} alignments, the information estimates are consistent across data sets. As expected, successive alignment procedures remove systematic variability and increase the information by comparable amounts, so that the maximal differences (between the minimal {\tt Y} alignment and the maximal {\tt XYT} alignment) are approximately 25\%, 21\%, 21\%, and 11\% for \emph{kn}, \emph{kr}, \emph{gt}, and \emph{hb}, respectively, when averaged across datasets.

\subsubsection{Merging data from different experiments}

To compute positional information carried by multiple genes from our data using the second Gaussian approximation, we need to measure $N$ mean profiles, $\bar{g}_i(x)$, and the $N\times N$ covariance matrix, $C_{ij}(x)$. Measuring individual gap gene expression profiles using, e.g., immunostaining, is a standard experimental technique in developmental biology. In contrast, estimating the covariance matrix, $C_{ij}(x)$, would require simultaneously labeling all $N$ gap genes in each embryo  using fluorescent probes of different colors. While simultaneous stainings of two genes are not unusual, it is not easy to scale the method up to more genes while maintaining a precise and quantitative readout. This is not a concern for the datasets analyzed in this paper, in which all gap genes were simultaneously recorded, but can be a concern when applying our method more generally. Therefore we present an estimation technique for inferring  a consistent $N\times N$ covariance matrix based on a collection of  embryos stained for different pairs of  genes. 

Estimating a joint covariance matrix from  pairwise staining  experiments is a non-trivial problem for two reasons. First,  each diagonal element of the covariance matrix, i.e. the variance of an individual gap gene, is measured in multiple experiments, but the obtained values might vary due to  statistical and systematic measurement errors. Second, true covariance matrices are positive definite, i.e. $\det(C(x))>0$,  a property that is not guaranteed by naively filling in different terms of the matrix by computing them across sets of embryos collected in different experiments. This is a consequence of small sampling errors that can strongly influence the determinant of the matrix. We therefore need a principled way to find a single best and valid covariance matrix from multiple partial observations, a problem that has had considerable history in statistics and finance (see, e.g., \cite{stein56,newey}).

We start by considering a number $\mathcal{N}_{ij}$ of embryos that have been co-stained for the pair of gap genes $(i, j)$. Let the full dataset consist  of all such pairwise stainings: for N gap genes, this is a total of ${N \choose 2}$ pairwise experiments,  where $i,j=1,\dots,N$ and $i<j$. Thus, in the case of the four major gap genes in \emph{Drosophila} embryos, \textit{kn}, \textit{kr},  \textit{gt} and \textit{hb}, the total number of recorded embryos is $\mathcal{N}=\sum_{(i,j)}\mathcal{N}_{ij}$, where the sum is across all six pairwise measurements: $(1,2),(1,3),(1,4),(2,3),(2,4),(3,4)$. These pairwise measurements give us estimates of the mean profile and the $2\times 2$ covariance matrices for each pair $(i,j)$:
\begin{eqnarray}
\hat{g}_i(x) &=& \frac{1}{\mathcal{N}_{ij}}\sum_{\mu=1}^{\mathcal{N}_{ij}}g_i^{(\mu)}(x), \label{ex1} \\
\hat{C}_{ij}(x) &=& \frac{1}{\mathcal{N}_{ij}}\sum_{\mu=1}^{\mathcal{N}_{ij}}(g_i^{(\mu)}(x)-\hat{g}_i)(g_j^{(\mu)}(x)-\hat{g}_j). \label{ex2}
\end{eqnarray}

The index $\mu$ enumerates all the embryos recorded in a pairwise experiment $(i,j)$. For four gap genes and six pairwise experiments, we will get six $2\times 2$ partial covariance matrices $\hat{C}$, and $6 \times 2$ estimates of the mean profile $\hat{g}$. Our task is to find a single set of 4 mean profiles $\bar{g}_i(x)$, and a single $4\times 4$ covariance matrix $C_{ij}(x)$, that fits all pairwise experiments best. In the next paragraphs, we will show how this can be computed for the arbitrary case of $N$ gap genes. 

To infer a single set of mean profiles $\bar{g}_i(x)$ and a single consistent $N\times N$ covariance matrix $C_{ij}(x)$, we  use maximum likelihood inference. We  assume that at each position $x$  our data is generated by a single $N$-dimensional Gaussian distribution of Eq.~(\ref{gaussian}) with unknown mean values and an unknown covariance matrix, which we would like to find, but we can only observe two of the mean values and a partial covariance in each experiment; the other variables are integrated over in the likelihood.

Following this reasoning, the log likelihood of the data for pairwise staining $(i,j)$ at position $x$ is
\begin{eqnarray}
 \mathcal{L}_{ij}&=&\frac{1}{\mathcal{N}_{ij}}\log \int \prod_{k\neq i,j} dg_k\; P(\{g_i\}|x) =\\
&=&\frac{1}{\mathcal{N}_{ij}}\ln\prod_{\mu=1}^{\mathcal{N}_{ij}}\frac{1}{2\pi\sqrt{C_{ii}C_{jj}-C^2_{ij}}} \times \nonumber\\
&\times&\exp\!\left[\!-\frac{1}{2}\frac{C_{jj}(g_i^{(\mu)} \!\!- \bar{g}_i)^2 \!+\! C_{ii}(g_j^{(\mu)} \!\!-\bar{g}_j)^2\!-\!2C_{ij}g_i^{(\mu)}g_j^{(\mu)}}{C_{ii}C_{jj}-C^2_{ij}}\!\right]\!\!, \nonumber
\end{eqnarray}
where the log likelihood $\mathcal{L}$, as well as the mean profiles, covariance elements and the measurements all depend on $x$. 

Since all pairwise experiments are independent measurements, the total likelihood $\mathcal{L}_{\rm tot}(x)$ at a given position $x$ is the sum of the individual likelihoods
\begin{equation}
\mathcal{L}_{\rm tot}(x)=\sum_{(i,j)}\mathcal{L}_{ij}(x)
\end{equation}
After some algebraic manipulation, the total log likelihood can  be written as:
\begin{eqnarray}
\lefteqn { \mathcal{L}_{\rm tot}(x)\!=\!-\!\!\sum_{(i,j)}  \!\ln(2\pi)\!+\!\ln\left(C_{ii}(x)C_{jj}(x)\!-\!C^2_{ij}(x)\right) } \\
&& +C_{jj}(x)\frac{\hat{C}_{ij}(x)-2\bar{g}_i(x)\hat{g}_i(x)+\bar{g}_i^2(x)}{C_{ii}(x)C_{jj}(x)-C^2_{ij}(x)} \nonumber \\
&& +C_{ii}(x)\frac{\hat{C}_{jj}(x)-2\bar{g}_j(x)\hat{g}_j(x)+\bar{g}_j^2(x)}{C_{ii}(x)C_{jj}(x)-C^2_{ij}(x)} \nonumber \\
&& +2C_{ij}(x)\frac{\hat{C}_{ij}(x)\!-\!\hat{g}_i(x)\bar{g}_j(x)\!-\!\hat{g}_j(x)\bar{g}_i(x)\!+\!\bar{g}_i(x)\bar{g}_j(x)}{C_{ii}(x)C_{jj}(x)\!-\!C^2_{ij}(x)},\nonumber
\end{eqnarray}
where $\hat{g}_i(x)$ and $\hat{C}_{ij}(x)$ are the experimentally determined profiles and covariance elements defined in Eqs.~(\ref{ex1},\ref{ex2}).
For each position $x$, we search for $\bar{g}_i(x)$ and $C_{ij}(x)$ that maximize $\mathcal{L}_{\rm tot}(x)$. Before proceeding, however, we have to guarantee that the search can only take place in the space of positive semi-definite matrices $C_{ij}(x)$ (i.e. $\mathrm{det}\,\mathbf{C}\geq 0$).  We enforce this constraint by spectrally decomposing $C_{ij}$ and parametrizing it in its eigensystem \cite{Pinheiro07}. To this end, we write
\begin{equation}
\mathbf{C}(x)=\mathbf{P}\mathbf{D}\mathbf{P}^{\mathrm{T}}, \label{crep}
\end{equation}
where  $\mathbf{D}$ is a diagonal matrix parametrized with the variables $\alpha_1,\dots,\alpha_N$ that determine the diagonal elements in $\mathbf{D}$, i.e. $D_{ii}=\exp(\alpha_i)$. The orthonormal matrix $\mathbf{P}$ is decomposed as a product of the $N(N-1)/2$ rotation matrices in $N$ dimensions:
\begin{equation}
\mathbf{P}=\prod_{k=1}^{N(N-1)/2}\mathbf{R}_k(\varphi_k),\label{pmatrix}
\end{equation}
where $\mathbf{R}_k(\varphi_k)$ is a rotation matrix that can be written as:
{\setlength{\arraycolsep}{2pt}
\begin{equation}
\mathbf{R}_k(\varphi_k) \! = \! \!  
\begin{pmatrix}
1 &  &  & & & & & & & \\
 &  &  \ddots &  & & & & & & \\
 &  &  &1 \!\!& & & & & &  \\
 &  &  & & \cos(\varphi_k)& 0 & \cdots & 0 & -\sin(\varphi_k) &  \\
 &  &  & & 0 & 1& & & 0 &  \\
 &  &  & &  \vdots & & \ddots & & \vdots &  \\
 &  &  & &  0 & & & 1 & 0 &  \\
 &  &  & & \sin(\varphi_k) & 0 & \cdots & 0 & \cos(\varphi_k)& \\
 &  &  & & & & & & & 1 \\   
\end{pmatrix} \!\! . \label{rott}
\end{equation}
}
 
In the representation given by Eq~(\ref{crep}), the likelihood is a function of $N$ values $\bar{g}_i$, $N$ parameters $\alpha_i$, and $N(N-1)/2$ angles $\varphi_k$ at each $x$. In case of 4 gap genes, this is a total of 14 parameters that need to be computed by maximizing $\mathcal{L}_{\rm tot}(x)$ at each location $x$.

There is no guarantee that there is a unique minimum for the log likelihood; moreover, there could exist  sets of covariance matrices that all lead to essentially the same value for $\mathcal{L}_{\rm tot}(x)$, or, even more dangerously, the maximum likelihood solution could favor matrices with vanishing determinants, especially when estimating from a small number of samples. To address these issues, we regularize the problem by replacing $\mathbf{D}\rightarrow \mathbf{D}+\lambda\mathbf{I}$, where $\mathbf{I}$ is the identity matrix and $\lambda$ is the regularization parameter. Larger values of $\lambda$ will favor more ``spherical'' distributions, while small values will allow distributions that can be very squeezed in some directions. We thus maximize $\mathcal{L}_{\rm tot}(x;\lambda)$ to find the best set of parameters given the value of the regularizer (which is assumed to be the same for every $x$).  To set $\lambda$ we use cross-validation: the maximum likelihood fit is performed for various choices of $\lambda$ not over all available data (embryos), but only over a \emph{training} subset. The remaining embryos constitute a \emph{test} subset. Parameter fits for different $\lambda$ obtained on the training data can be assessed and compared by evaluating their likelihood over test data, and selecting the value of $\lambda$ that maximizes the model likelihood over the testing set.

To compute $\bar{g}_i(x)$ and $C_{ij}(x)$, we initialize $\bar{g}_i(x)$ to the mean profile across all pairwise experiments $(i,j)$; we initialize $\alpha_i$ to the mean of the log of the diagonal terms $\hat{C}_{ii}$; and we initialize all rotation angles $\varphi_k=0$. Nelder-Mead simplex method is used to maximize $\mathcal{L}_{\rm tot}(x)$ \cite{Lagarias98}. Finally, we compute $\mathbf{C}(x)$ from $\alpha_i$ and $\varphi_k$ using Eqs~(\ref{crep},\ref{pmatrix},\ref{rott}).

\begin{figure}
\centering
\includegraphics[width = 3.3in]{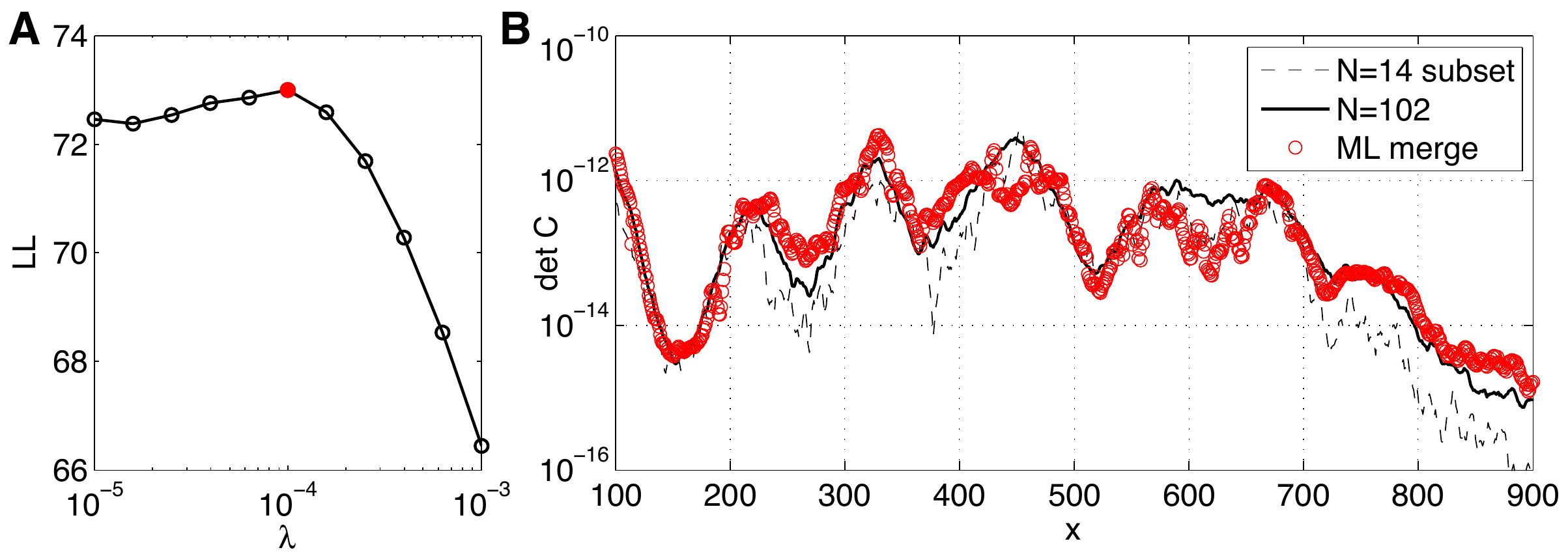}
\caption{\label{f9} {\bf Merging data sets using maximum likelihood reconstruction.} Dataset D  (102 embryos) is used to simulate 6 pairwise staining experiments with 14 unique embryos per experiment. {\bf A)} The log likelihood of the merged model on 18 test embryos (not used in model fitting), as a function of regularization parameter $\lambda$; $\lambda=10^{-4}$ is the optimal choice for this synthetic dataset. {\bf B)} The comparison of the determinant of the $4\times 4$ inferred covariance matrix (red) as a function of position $x$, compared to the determinant of the full dataset (black solid line) or the determinant of a subset of 14 embryos (black dashed line), where small sample effects are noticeable. }
\end{figure}

Because we have a quadruple stain, we can use it as the ground truth against which we can assess our method. We can generate ``synthetic'' experiments by taking 102 recorded embryos of Dataset D, partition that data into 6 disjoint subsets of 14 embryos each (while retaining 18 embryos for validation), and pretend that in each subset we only could record from one of the six possible pairs of gap genes; that is, in every subset for a pair of genes $(i,j)$, we measured the mean expression profiles $\bar{g}_i(x), \bar{g}_j(x)$ and the $2\times 2$ covariance matrix $C_{ij}(x)$, the inputs to the maximum likelihood merging procedure. The inferred $4\times 4$ covariance matrix can be compared to the ground truth covariance matrix estimated from the complete Dataset D data. The results of this procedure are shown in Fig~\ref{f9}, demonstrating that pairwise measurements can be merged into a consistent covariance matrix, whose determinant matches well the real determinant (determinants are compared because they are very sensitive to the inferred parameters and enter directly into the expressions for positional information and positional error). Note that filling in the covariance matrix naively, or skipping the regularization, would result in either non-positive-definite matrices, or matrices whose determinants are close to singular at multiple values of position $x$. We hope that the presented maximum likelihood merging procedure will allow our framework to be applied to model systems where simultaneous gap gene measurements are hard to obtain, but pairwise stains are feasible.

\subsubsection{Monte Carlo integration of mutual information}

Another technical challenge in applying the second Gaussian approximation for positional information to three or more gap genes lies in computing the entropy of the distribution of expression levels, $S[P_g(\{g_i\})]$. This is a Gaussian mixture obtained by integrating $P(\{g_i\}|x)$ over all $x$ as prescribed by Eq~(\ref{int2nd}). In one or two dimensions one can evaluate this integral numerically in a straightforward fashion by partitioning the integration domain into a grid with fine spacing $\Delta$ in each dimension, evaluating the conditional distribution $P(\{g_i\}|x)$ on the grid for each $x$, and averaging the results over $x$ to get $P_g(\{g\})$, from which the entropy can be computed using Eq~(\ref{sdef}). Unfortunately, for three genes or more this is  infeasible because of the curse of dimensionality for any reasonably fine-grained partition. 

To address this problem we make use of the fact that over most of the integration domain $P_g(\{g_i\})$ is very small if the variability over the embryos is small. This means that most of the probability weight is concentrated in the small volume around the path traced out in $N$-dimensional space by the mean gene expression trajectory, $\bar{g}_i(x)$, as $x$ changes from 0 to 1. We  designed a method that partitions the whole integration domain adaptively into volume elements such that the total probability weight in every box is approximately the same, ensuring fine partition in  regions where the probability weight is concentrated, while simultaneously only using a tractable number of partitions.

\begin{figure}
\centering
\includegraphics[width = 3.3in]{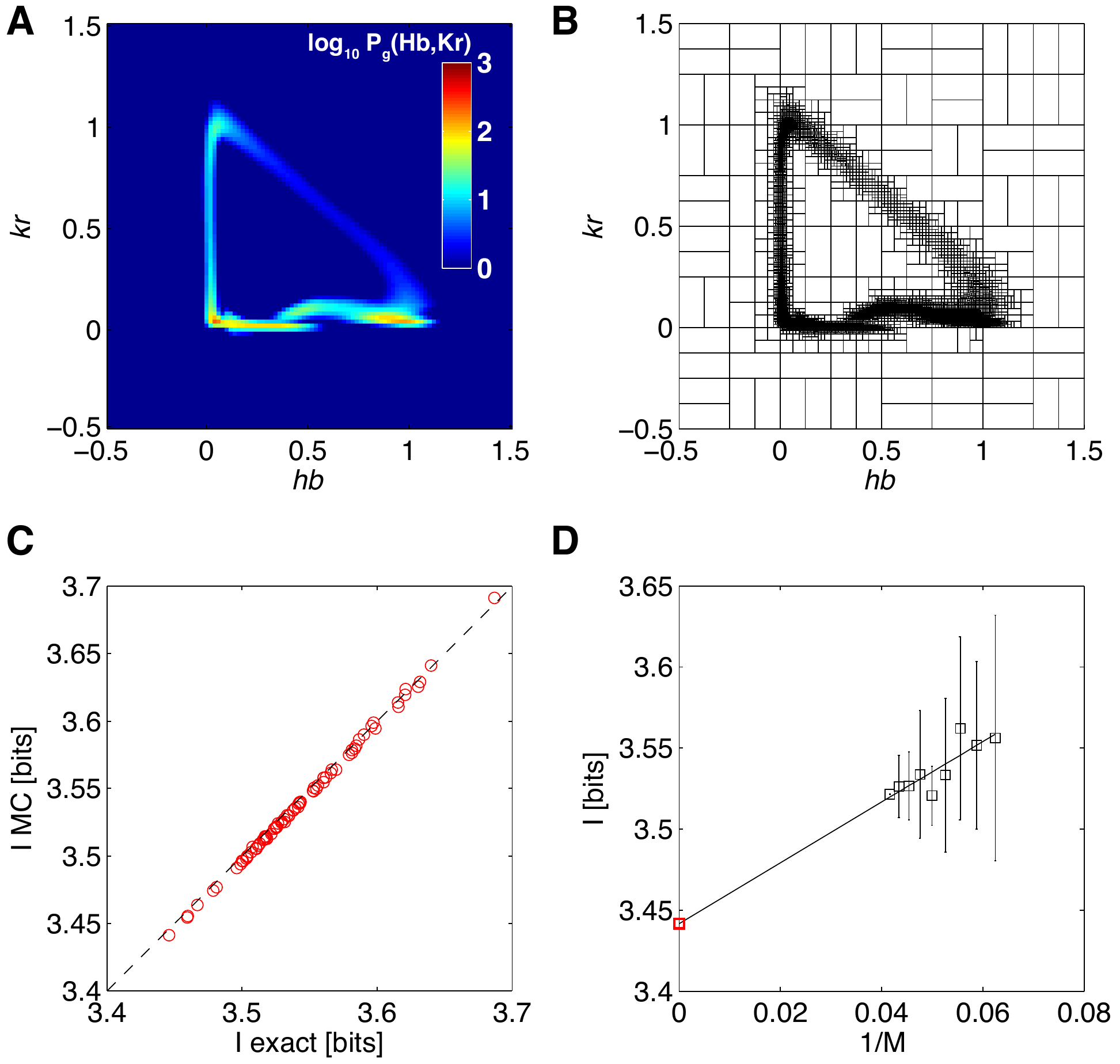}
\caption{\label{f10} {\bf Monte Carlo integration of information for a pair of genes.} {\bf A)} The (log) distribution of gene expression levels, $P_g(hb,kr)$, obtained by numerically averaging the conditional Gaussian distributions for the pair over all $x$, using {\tt Y} aligned  Dataset A, and including 10-90\% egg length segment. The distribution is evaluated over a grid of $100\times 100$ points, normalized such that $\int_{hb,kr=-0.5}^{1.5} dhb\;dkr P(hb,kr)=1$ and plotted on the log scale (color bar at right). For two genes this explicit construction of $P_g$ is tractable and can be used to evaluate the total entropy, $S[P_g]$. {\bf B)} As an alternative, one can use the Monte Carlo procedure outlined in the text which uses adaptive partitioning of the domain, shown here. The boxes, here $10^4$, are dense where the distribution contains a lot of weight. {\bf C)} The comparison between MC evaluated positional information carried by the \emph{hb/kr} pair and the exact numerical calculation as in A, over different random subsets of $m=16,\dots,24$ embryos of Dataset A, with 10 random subsets at every $m$. The MC integration underestimates the true value by $0.1\%$ on average, with a relative scatter of  $4\cdot 10^{-4}$. {\bf D)} To debias the estimate of the information due to the finite sample size used to estimate the covariance matrix, a SGA extrapolation to infinite data size, $m\rightarrow \infty$, is performed on the estimates from C, to yield a final estimate of $I(\{hb,kr\};x)=3.44\pm0.02$ bits. }
\end{figure}

\begin{figure}
\centering
\includegraphics[width = 2.5in]{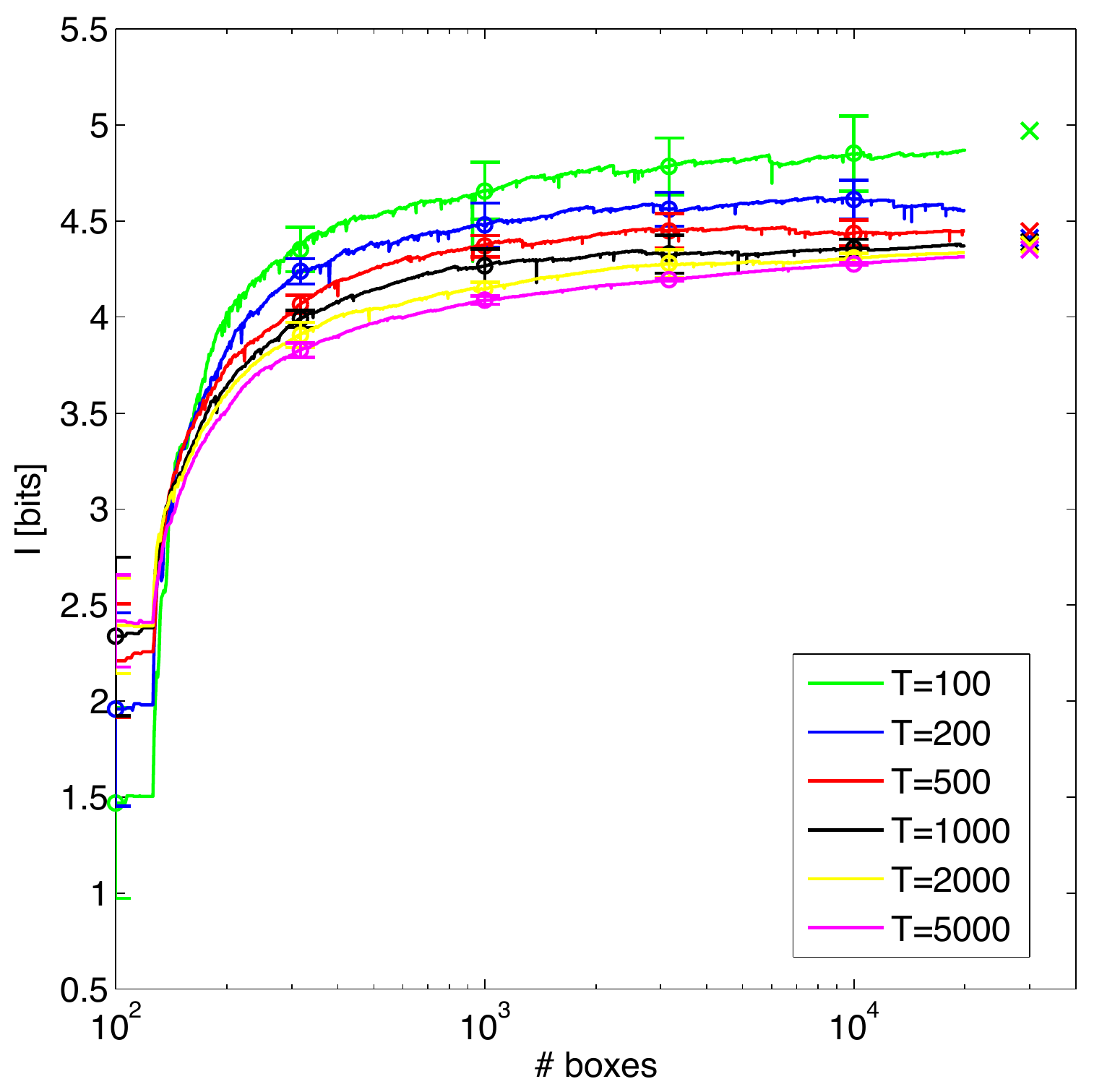}
\caption{\label{f11} {\bf Monte Carlo integration of mutual information for the gap gene quadruplet.} The estimation is performed with $\mathcal{N}=24$ Dataset A embryos over 10-90\% egg length segment using {\tt Y} alignment but \emph{without} correcting for the  finite number of embryos (i.e., the empirical covariance matrix is taken to be the ground truth for this analysis). For each choice of the number of sampling points per box, $T$, 10 independent MC estimations were run; shown is the mean convergence of positional information with error bars showing $1$-std scatter over the 10 estimator runs. As $T$ is increased, the estimator variance decreases, because the estimation within each box is precise enough to lead to a reproducible order of recursive box partitioning, eliminating large jumps in the estimate seen for small $T$. The information can then be linearly regressed against the inverse number of boxes for last 2000 partitions, and extrapolated to an infinitely fine partition; for each $T$, these extrapolations are shown at right (crosses).}
\end{figure}

The following algorithm was used to compute the total entropy, $S[P_g(\{g_i\})]$:
\begin{enumerate}
\item The whole domain for $\{g_i\}$ is recursively divided into boxes such that no box contains more than 1 percent of the total domain volume.
\item For each box $i$ with volume $\omega_i$, we use Monte Carlo sampling to randomly select $t=1,\dots,T$ points $\mathbf{g}^t$ in the box  and approximate the weight of the box $i$ as $\pi(i|x) = \omega_i T^{-1} \sum_{t=1}^T P(\mathbf{g}^t|x)$; we explore different choices for $T$ in Fig~\ref{f11}.
\item Analogously, we evaluate the approximate total weight of each box $\pi(i)$, by pooling Monte Carlo sampled points across all $x$.
\item $\pi(i|x)$ and $\pi(i)$ are renormalized to ensure $\sum_i \pi(i|x)=1$ for every $x$ and $\sum_i \pi(i)=1$.
\item The conditional and total entropies are computed as $S_{\rm noise}=-\langle \sum_i\pi(i|x)\log_2\pi(i|x)\rangle_x$ and $S_\mathrm{\rm tot}=-\sum_i\pi(i)\log_2\pi(i)$.
\item The positional information is estimated as $I(\{g_i\};x)=S_\mathrm{tot}-S_\mathrm{noise}$.
\item The box $i^*=\mathrm{argmax}_i \pi(i)$ with the highest probability weight $\pi(i^*)$ is split into two smaller boxes of equal volume $\omega(i^*)/2$, and the estimation procedure is repeated by returning to step 2. Additional Monte Carlo sampling only needs to be done within the newly split box; for the other boxes old samples can be reused.
\item The algorithm terminates when the positional information achieves desired convergence, or at a preset number of box partitions.
\end{enumerate}

 Figures~\ref{f10}A and \ref{f10}B show how the algorithm works for a pair genes ($\{hb,kr\}$) for which the joint probability distribution is easy to visualize.  The resulting adaptive partition is spatially highly refined where the distribution $P_g(\{hb,kr\})$ has a lot of weight and remains coarse elsewhere. For $T=100$ and adaptive partitioning of the domain into $10^4$ boxes, the relative difference between the MC estimate of the information and the evaluation over an uniform grid is $\sim 10^{-3}$; this excellent match is shown in Fig~\ref{f10}C. To get the final SGA estimate of positional information that the two genes jointly carry about position, extrapolation to infinite data size is performed in Fig~\ref{f10}D.

Figure~\ref{f11} shows the MC estimate of the positional information carried by the quadruplet of gap genes, and its dependence on the parameter $T$ (the number of MC samples per box per position) and the refinement of the adaptive partition. When $T$ is too small (e.g., $T=100$) we obtain a biased estimate of the information, presumably because the  evaluation of the distribution over boxes is poor, leading to suboptimal recursive partitioning. When $T$ is increased to $T\geq 200$, the estimates for different $T$ start converging as the number of adaptive boxes is increased, and the final estimates (after extrapolation to an infinitely fine partition) for different $T$ agree to within a few percent.

%
%
%
%

%
%
% RESULTS
%
%
%
%
%
\subsection{Application to the  \emph{Drosophila} gap gene system}

In this section we apply the information estimation methods developed above to measure the information content in 1) individual gap genes, 2) pairs of gap genes, and 3) the full set of four gap genes. The latter we will refer to as the {\it gene quadruplet}. To make our analyses consistent and comparable with previous work \cite{PNASPI}, we focus only on dorsal gene expression profiles in the  central 10--90\% segment of the embryos in the 38--48 min time class, using the minimal ({\tt Y}) data alignment. We comment on the above-mentioned alternative alignment methods below.

\subsubsection{Information in single gap genes and gap gene pairs}
Information carried by single genes is reported in Table~\ref{InfoTable}. The values are estimated using the direct method which agrees closely with the alternative estimation methods (FGA, SGA; cf. Fig.~\ref{f7}). Measured across four independent experiments, the information values are consistent to within the estimated error bars, indicating that not only embryo-to-embryo experimental variability can be brought under control as described in Ref~\cite{MSB}, but also that experiment-to-experiment variability is small.

\begin{table} 
\centering \begin{tabular}{| c | c | c |  c | c | c | c | c | c | c | c | c | c |} 
\hline\hline 
& \multicolumn{3}{c|}{\textit{kn}} & \multicolumn{3}{c|}{\textit{kr}} & \multicolumn{3}{c|}{\textit{gt}} & \multicolumn{3}{c|}{\textit{hb}}\\
\hline 
 A & \multicolumn{3}{c|}{$1.82\pm.06$} &\multicolumn{3}{c|}{$1.94\pm.07$}&\multicolumn{3}{c|}{$1.81\pm.06$}&\multicolumn{3}{c|}{$2.26\pm.04$} \\ 
 B & \multicolumn{3}{c|}{$1.81\pm.04$} &\multicolumn{3}{c|}{$1.93\pm.06$}&\multicolumn{3}{c|}{$1.79\pm.05$}&\multicolumn{3}{c|}{$2.29\pm.05$} \\ 
 C & \multicolumn{3}{c|}{$1.88\pm.07$} &\multicolumn{3}{c|}{$2.04\pm.05$}&\multicolumn{3}{c|}{$1.84\pm.05$}&\multicolumn{3}{c|}{$2.19\pm.05$} \\ 
 D & \multicolumn{3}{c|}{$1.79\pm.04$} &\multicolumn{3}{c|}{$1.94\pm.04$}&\multicolumn{3}{c|}{$1.91\pm.03$}&\multicolumn{3}{c|}{$2.21\pm.03$} \\ 
\hline
Mean & \multicolumn{3}{c|}{$1.83\pm.04$} &\multicolumn{3}{c|}{$1.96\pm.05$}&\multicolumn{3}{c|}{$1.84\pm.05$}&\multicolumn{3}{c|}{$2.24\pm.05$}  \\
\hline 
\hline 

\hline 
\end{tabular} 
\caption{\label{InfoTable} {\bf Information (in bits) carried by single genes.} For each dataset, we report the direct estimate of positional information for the  10-90\% egg length segment, after {\tt Y} alignment. Error bars are the estimation errors. The last row contains the (mean $\pm$ std) over datasets. } 
\end{table}

\begin{figure}
\centering
\includegraphics[width = 3in]{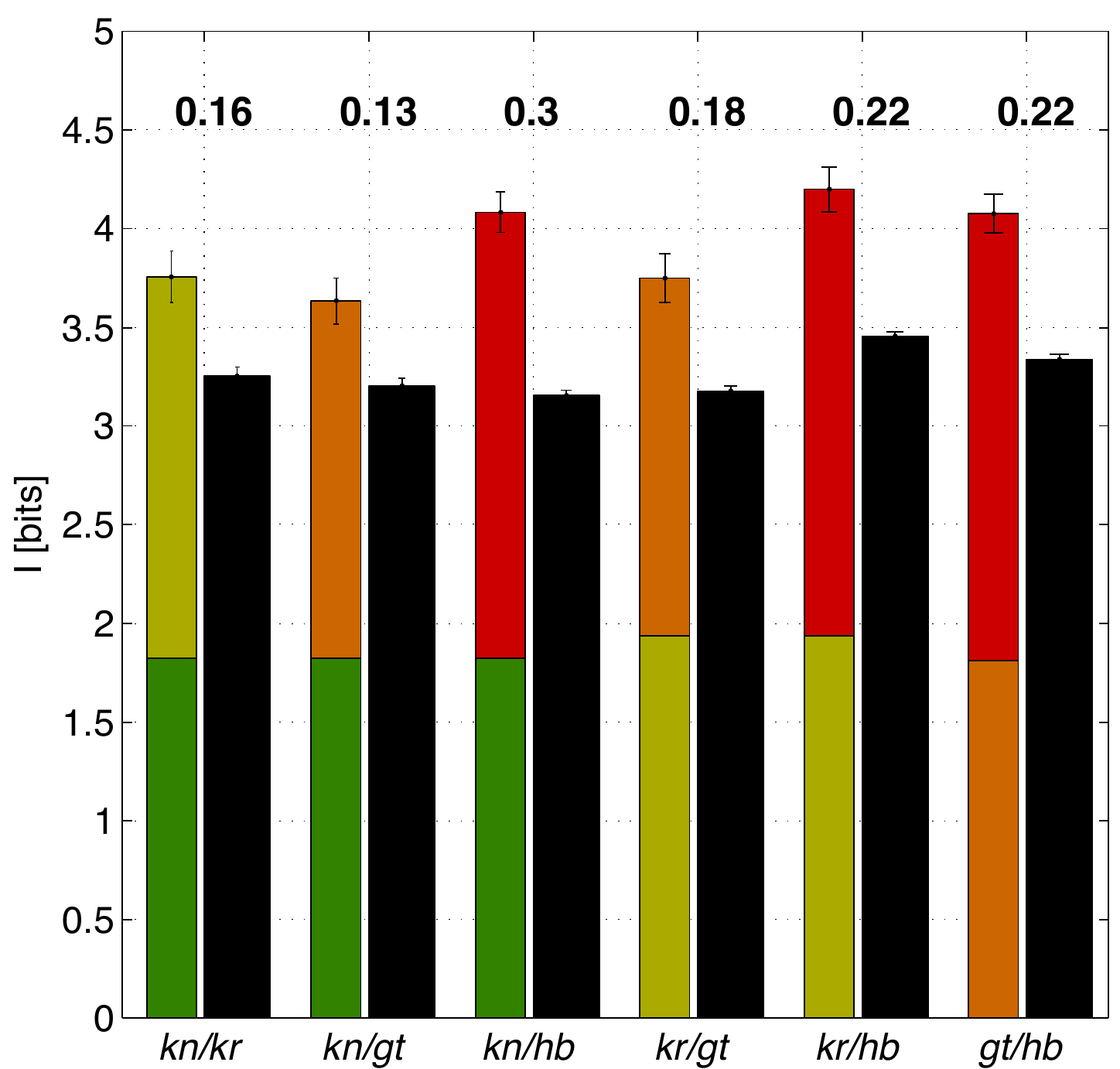}
\caption{\label{f13} {\bf Positional information in 10-90\% egg length segment carried by pairs of gap genes and their redundancy.} For each gene pair (horizontal axis) the first bar represents the sum of individual positional informations (color coded as in Fig.~\ref{f5}), using direct estimate after {\tt Y} alignment of Dataset A. The second (black) row represents the positional information carried jointly by the gene pair, estimated using SGA with direct numerical evaluation of the integrals.  All pairs are redundant, with fractional redundancy $R=\left(I(g_1;x)+I(g_2;x)-I(\{g_1,g_2\};x)\right)/I(\{g_1,g_2\};x)$ reported above the bar. $R=1$ for a totally redundant pair, $R=0$ for a pair that codes for the position independently, and $R=-1$ for a totally synergistic pair where individual genes carry zero information about position. }
\end{figure}

Single genes carry substantially more than 1 bit of positional information, sometimes even exceeding 2 bits. Thus each gap gene alone provides more information about position than what would be conveyed by the hypothesized ``decision threshold'' that separates the embryo into ``on'' and ``off'' expression domains. But this information is redundantly encoded: the sum of individual positional informations for single genes is always greater than the jointly encoded information (Fig.~\ref{f13}), although redundancy  at the pairwise level is relatively small ($\sim\!20\%$). 
Such a low degree of redundancy is expected because gap genes are often expressed in complementary regions of the embryo in non-trivial combinations and will thus mostly convey new information about position. Unlike in our toy examples in the Introduction, real gene expression levels are continuous and noisy, and some degree of redundancy might be useful in mitigating the effects of noise, as has been shown for other biological information processing systems \cite{GTneurons}. Overall, with the addition of the second gene positional information increases well above the 0.5 bits that would be expected theoretically for fully redundant gene profiles. The increase is also larger than the theoretical maximum for non-redundant (but non-interacting) genes, where the increase for the second gap gene would be limited to 1 bit \cite{Tkacik09}. The ability to generate non-monotonic profiles of gene expression (e.g.~bumps) is therefore crucial for high information transmissions achieved in the \emph{Drosophila} gap gene network \cite{Walczak10}.

\subsubsection{Information  in the gap gene quadruplet}

\begin{figure*}
\centering
\includegraphics[width = 6in]{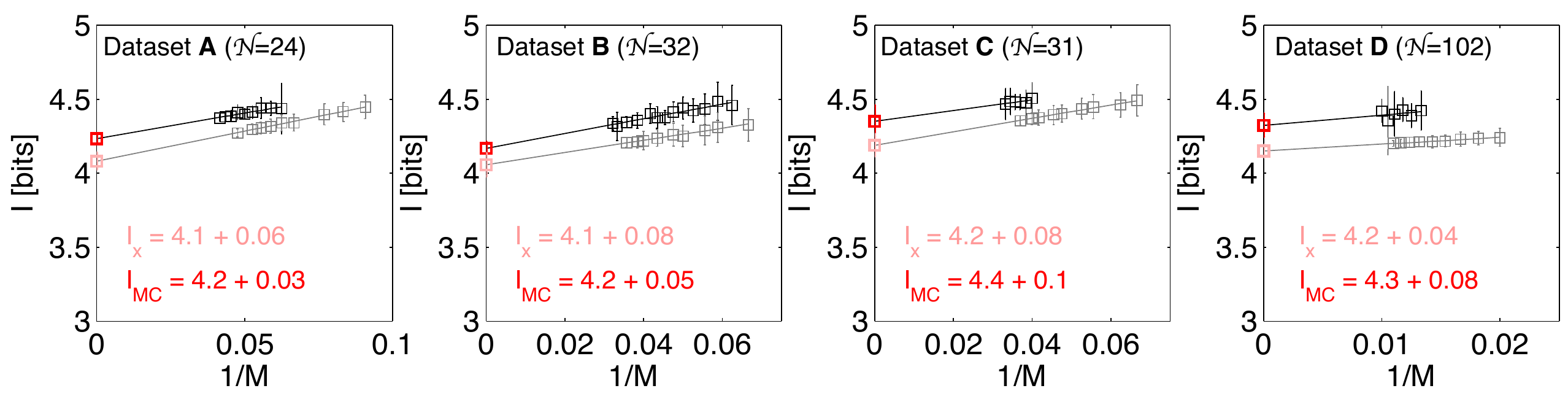}
\caption{\label{f16} {\bf Positional information carried by the quadruplet, $\{kn,kr,gt,hb\}$, of gap genes.} Each panel shows the positional information estimation for a separate dataset (A, B, C, D, respectively) in the 10-90\% egg length segment, using {\tt Y} data alignment method. Shown is the extrapolation to infinite data limit ($M\rightarrow \infty$) for SGA estimates using Monte Carlo integration (black plot symbols and dark red extrapolated value in bits), as well as for estimation of positional information from positional error (gray plot symbols and bright red extrapolated value in bits). For MC estimation, 25 subsets of embryos were analyzed per subset size. For estimation from positional error, 100 estimates at each subsample size (data fractions $f=0.5, 0.55, \dots, 0.85, 0.9$) were performed. In the MC estimation, where MC sampling contributes to the error on the extrapolated information value, the error on the final estimate is taken to be the std of the estimates over 25 subsets at the highest data fraction (smallest $1/M$). For estimation from positional error that doesn't use a stochastic estimation procedure, the error estimate on the final extrapolation is the variance due to small number of samples, estimated as $2^{-1/2}$ times the std over 100 estimates at half the data fraction.  }
\end{figure*}

How much information about position is carried by the four gap genes together? Figure~\ref{f16} shows the estimation of positional information using MC integration for all four datasets. 
The average positional information across four datasets is $I=4.3\pm 0.07$ bits, and this value is highly consistent across the datasets. Notably, the information content in the quadruplet displays a high degree of redundancy: the sum of single-gene positional information values is substantially larger than the information carried by the quadruplet, with the fractional redundancy 
\begin{equation}
R=\frac{\sum_{i=1}^4 I(g_i;x)-I(\{g_i\};x)}{I(\{g_i\};x)}
\end{equation}
being $R=0.84$. Possible implications of such a redundant representation are addressed in the Discussion.

To assess the importance of correlations in the expression profiles and their contribution to the total information, we start with the covariance matrices $C_{ij}(x)$ of the Dataset A, which we artificially diagonalize  by setting the off-diagonal elements to zero at every $x$. This manipulation destroys the correlations between the genes, making them conditionally independent (mechanistically, off-diagonal elements in the covariance matrix could arise as a signature in gene expression noise of the gap gene cross interactions). With these matrices in hand we performed the  information estimation using Monte Carlo integration, analogous to the analysis in Fig~\ref{f16}. Surprisingly, we find a minimal increase in information from $I=4.2\pm 0.03$ bits to $I=4.4\pm 0.02$ bits when the correlations are removed. This indicates that gap gene cross interactions, which are hypothesized to be important for shaping the mean expression profiles by generating bump-like (as opposed to step-like) spatial patterns, play a minor role in reshaping the gene expression variability, at least insofar as that influences the total positional information.

We also evaluated the importance of the alignment procedure for estimating total information. Again we use Dataset A to recompute the MC information estimates with  {\tt YT}, {\tt XY}, and {\tt XYT} alignments, and compare these with the {\tt Y} alignment procedure used in Fig~\ref{f16}. The positional information encoded by the quadruplet is $I=4.3\pm 0.03$ (for {\tt YT}), $I=4.7\pm 0.06$ (for {\tt XY}), and $I=4.8\pm 0.06$ (for {\tt XYT}), respectively. This shows that the temporal ({\tt T}) alignment does not change the information much, probably because our stringent initial selection cutoff on the depth of the membrane furrow canal has picked out the profiles that are sufficiently localized in time around their stable shapes. In contrast, {\tt X} alignment emerges as important, generating an extra half bit of positional information. This suggests that either our determination of the AP axis used to assign a coordinate to every gene expression has a random embryo-to-embryo error (which is unlikely given the visual inspection of microscopy images and extracted AP axes), or that the gap gene expression pattern is intrinsically variable in that it shifts rigidly from embryo to embryo relative to the egg boundary. This would imply that correlated readout errors that the nuclei might make are less harmful than uncorrelated errors. If two neighboring nuclei are positioned both towards the left or both towards the right of the ``true'' pattern with some positional error $\sigma_x$, this is very different from \emph{each} of the nuclei randomly perturbing its position with noise of magnitude $\sigma_x$: in the first case, the rank order of the nuclear identities is preserved, while in the second case it needn't be, possibly leading to a spatial mixing of the cell fates detrimental to the development process.

\subsubsection{Decoding information and the resulting positional error}

\begin{figure}
\centering
\includegraphics[width = 3.3in]{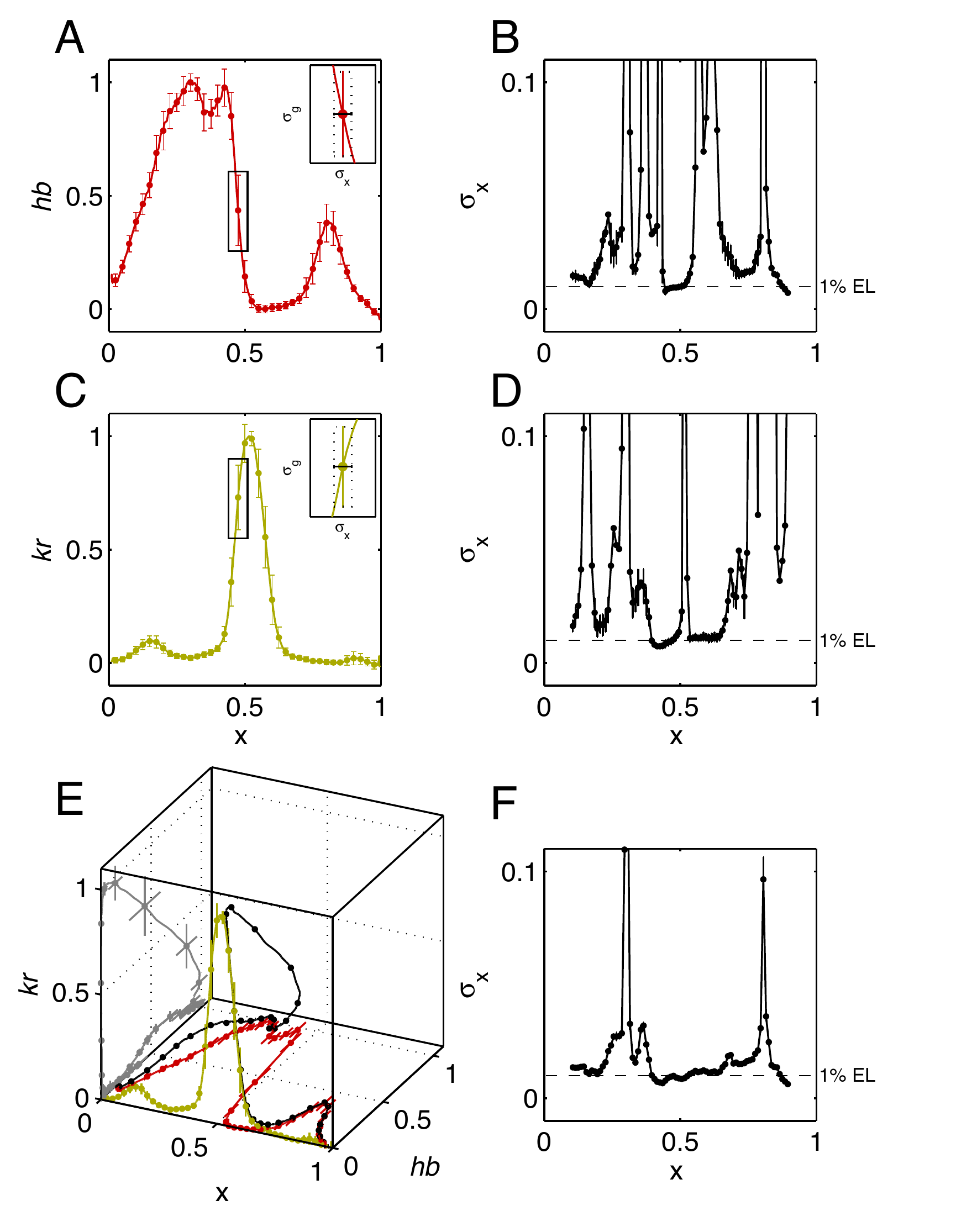}
\caption{\label{f12} {\bf Positional error for a single gene and a pair of genes.} {\textbf{A)}} The mean \textit{hb} profile and standard deviation across $\mathcal{N}=24$ embryos of Dataset A with {\tt Y} alignment, as a function of fractional egg length. The inset shows a close-up of the transition region with the positional error $\sigma_x$  determined geometrically from the mean $\bar{g}(x)$ and the standard deviation $\sigma_g$ of the profiles, as explained in the main text.  {\textbf{B)}} Positional error for \textit{hb} as a function of position, computed using Eq~(\ref{sigmax1}). The dashed line is a reference for $\sigma_x=0.01$ or 1\% EL. Error bars are obtained by bootstrapping 10 times over $\mathcal{N}/2$ embryo subsets. {\textbf{C, D)}} Plots analogous to A, B for \emph{kr}. {\textbf{E)}} Three-dimensional representation of \textit{hb} and \textit{kr} profiles from A and C as a function of  position $x$. The mean and standard deviations of the gene expression levels are shown  in the $\{hb,kr\}$ plane  in gray curve with errorbars; cf. the joint distribution of \emph{hb} and \emph{kr} expression levels in Fig~\ref{f11}A. The black curve that extends through the cube volume shows the average expression ``trajectory'' as a function of position $x$. {\textbf{F)}}  Positional error computed using Eq~(\ref{poserr}) for the $\{hb,kr\}$ pair, using their measured mean expression profiles and covariance. }
\end{figure}

Positional information is a single aggregate, or global, measure that quantifies the performance of the patterning system, but we can also ask about such performance position-by-position. A local measure based on estimation theory \cite{et} is the \emph{positional error}, which can be computed from the Fisher information of Eq~(\ref{fisher}).  Positional error is the smallest error by which the position can be determined from a local measurement of a set of noisy gene expression levels that is consistent with Bayes' optimal decision making. As a result, this quantity represents a bound on the precision of any cellular readout as well, and is therefore directly comparable to the measured precision of various positional markers.

We start by looking at a single gene $g$, for which we compute the positional error (for equally spaced positions along the AP axis) using Eq~(\ref{sigmax1}).
Figure~\ref{f12} illustrates this procedure geometrically for the case of \textit{hb} and \textit{kr}. The positional error at a given $x$ can be visualized as follows. Find the mean value of the expression profile at the desired $x$, $\bar{g}(x)$, and center on it a thin rectangle whose height is equal to the noise in gene expression, $\sigma_g(x)$ at that point. Increase the width of the rectangle and stop when the gap gene profile, $\bar{g}(x)$, intersects the rectangle in its corners, at which point the width of the rectangle will be roughly $\sigma_x(x)$ (insets to Fig~\ref{f12}A and C). Alternatively, this can be seen as propagating the variability in gene expression (vertical error bar) to an equivalent error in position (horizontal error bar) through the slope of the profile, $\bar{g}'(x)$. The positional error is plotted as a function of $x$ in Fig~\ref{f12}B and C for \emph{hb} and \emph{kr}, respectively.

%, 

An important advantage of using positional error as a local measure of precision in patterning is that it can be naturally generalized to quantify the precision of local position readout using more than one morphogen gradient. In Fig~\ref{f12}E and F we analyze the positional error given the joint readout of the $\{hb,kr\}$ pair.  The results illustrate that the optimal positional decoding performed with several  genes (e.g., $N=2$) at a given $x$ does not correspond to the positional error carried by the most informative gene at that position; the combined error can be smaller than the individual errors due to the noise averaging by the $N$ readouts, as well as due to the correlation structure in the variability of the $N$  profiles. For 2 genes, $\sigma_x$ can be interpreted geometrically as the AP distance between the positions of the intersection points of $\bar{g}(x)$ and a cylinder whose base is the ellipsoid in the $\{hb,kr\}$ plane such that $\sum_{i,j}(g_i(x)-\bar{g}_i)[C^{-1}(x)]_{ij}(g_j(x)-\bar{g}_j)\le1/4$.

\begin{figure}
\centering
\includegraphics[width = 3in]{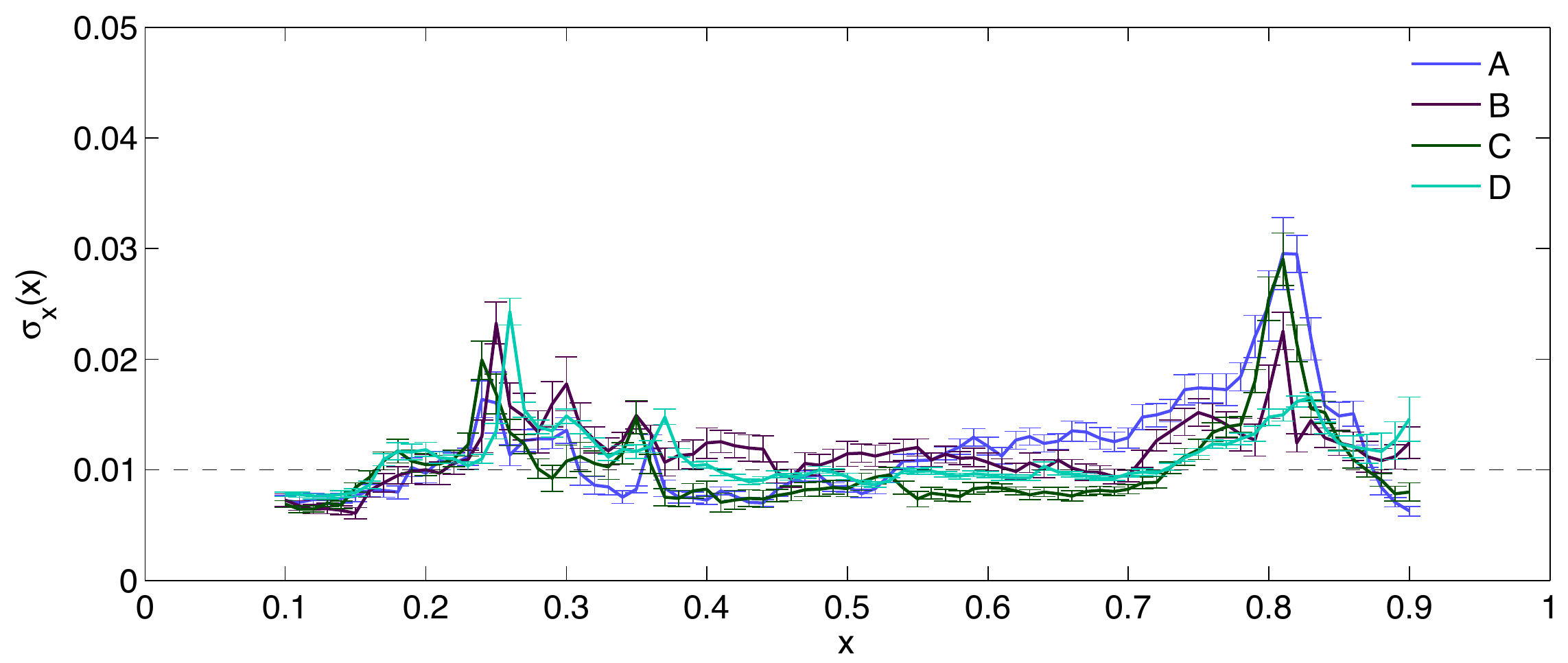}
\caption{\label{f14} {\bf Positional error for the gene quadruplet in 10-90\% egg length segment.} Positional error has been estimated using Eq~(\ref{poserr}) after {\tt Y} alignment, and extrapolated to large sample size. Error bars are bootstrap error estimates. Different colors denote different datasets (legend).}
\end{figure}

Figure \ref{f12} shows that by reading out single gap genes the nuclei can already achieve positional errors of less than 1.5\% egg length in specific regions of the embryo, yet fail to do so in other regions. In fact, positional error is seen to be smaller in the regions of high profile slope, where the variations in gene expression are reliably translated into variations in position. Conversely, the error formally diverges at the peaks and troughs of the profiles where small variations in gene expression cannot efficiently map to changes in position. This is consistent with the notion that, when noise is small enough, the positional information is encoded in the parts of the embryo where the morphogens have a large slope, rather than in domains where their expression level is roughly constant. Were the noise much higher, this conclusion would not hold---in that regime one could distinguish solely between, e.g., a domain of minimal and a domain of maximal expression, and the positional information would correspond better to the intuitive picture of gap genes that define ``on'' and ``off'' domains.

By considering pairs of genes, as in Fig~\ref{f12}E, F, the positional error can be made small across an increasing fraction of the AP axis. Finally, using the gap gene quadruplet simultaneously, the positional error can reach an average value of $\sim 1\%$, while never exceeding a few percent, as shown for all four datasets in Fig~\ref{f14}. This shows that the gap genes establish a convenient ``chemical coordinate system,'' in which it is in principle possible to position any feature along the entire length of the AP axis with roughly one percent precision; the uniform coverage of the AP axis is a sign of efficient encoding of positional information \cite{PNASPI}. In this regime of operation, the consequences of gene expression variability are truly small and we expect that the approximation to the positional information using positional error, given by Eq~(\ref{infox}), could hold.

\begin{figure}
\centering
\includegraphics[width = 3in]{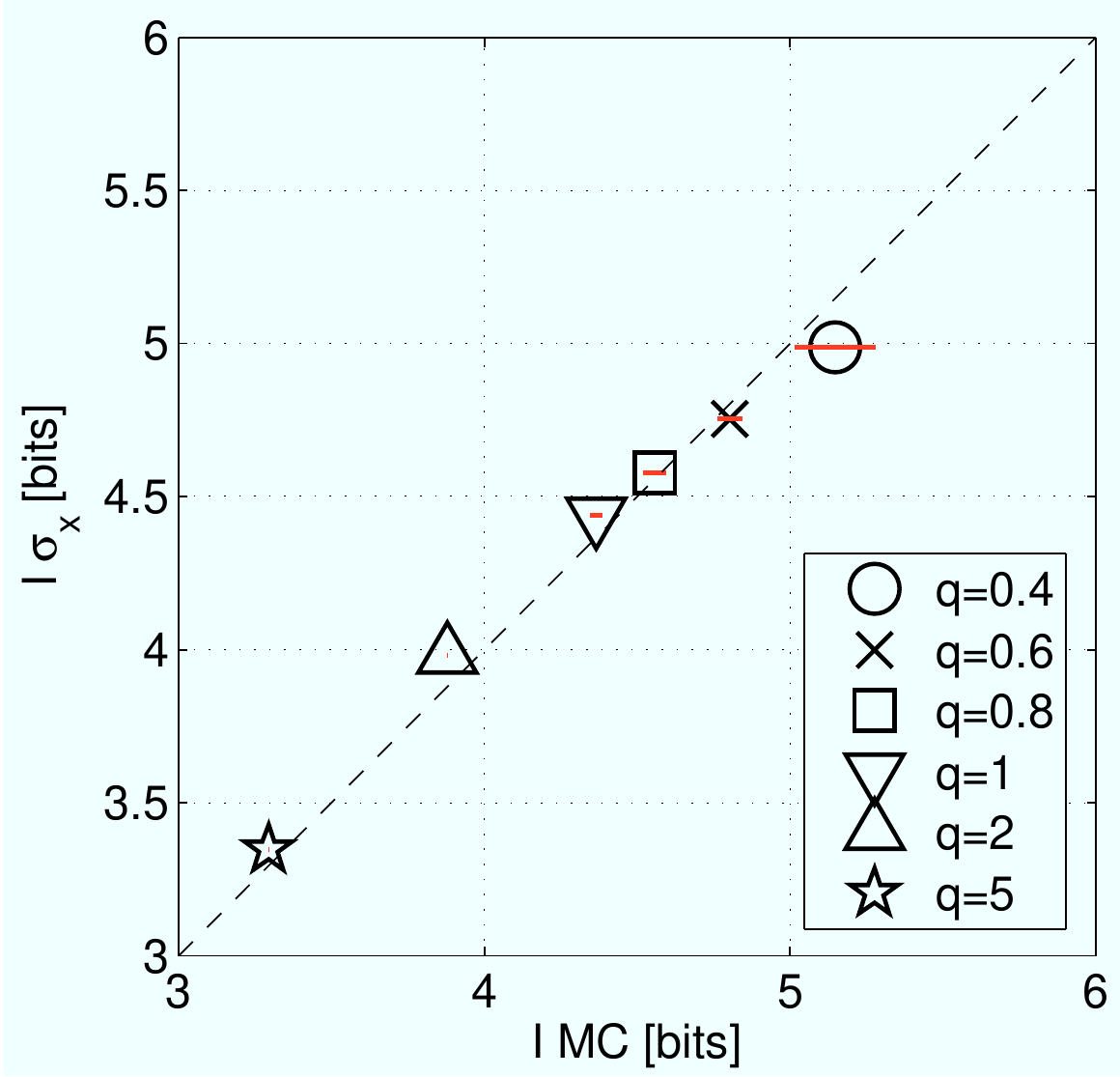}
\caption{\label{f15} {\bf The validity of the small noise approximation.} Synthetic data was generated by starting with {\tt Y} aligned profiles from Dataset A and proceeding by keeping the mean profiles as inferred from the data, forcing the covariance matrix to be diagonal (by zeroing out the off-diagonal terms), and multiplying the variances by a tunable factor $q$ (see legend).  This allowed us to manually tune the noise, with $q=1$ corresponding to measured variances in the data and $q<1$ ($q>1$) corresponding to decreasing (increasing) amount of variability. For each value of $q$, positional information was estimated from the positional error using Eq~(\ref{poserr}) (vertical axis), or using Monte Carlo integration (horizontal axis). Note that no small sample correction is needed here since the empirical estimates of mean profiles and covariance matrices are taken to be the ground truth. Horizontal error bars are std across 10 independent Monte Carlo integrations for every $q$. At low $q$,  the MC estimate has a large error bar, because the probability weight in the joint distribution is tightly concentrated around the mean gene expression levels and our estimation procedure (which does not assume this) initially has a hard time finding the region in expression space that it is supposed to recursively refine.}
\end{figure}

To systematically check if the gene quadruplet system really is within the small noise limit where  expressions of Eqs~(\ref{sigmax1},\ref{poserr},\ref{infox}) should hold, we perform the  analysis summarized in Fig~\ref{f15}. We systematically scale the measured gene variability in Dataset A up or down by a factor $q$ to generate synthetic datasets, and compare the positional information computed directly using MC integration on these synthetic data with the approximation of Eq~(\ref{infox}), computed using the positional error. Across the range of $q$ (extending from $\sqrt{0.4}\approx 0.63$ times the observed noise, to $\sqrt{5}\approx 2.24$ times the observed noise), the difference between both estimates of positional information is less than 3\%. For small $q$,  the information for the synthetic data (which is Gaussian by construction) should be equal to the information derived from positional error computed using the full expression for the Fisher information given by Eq~(\ref{fullFI}). In fact, Fig~\ref{f15} shows that simple approximations to the Fisher information leading to Eqs~(\ref{sigmax1},\ref{poserr}) are sufficient for a very good match. Even when the noise is increased for $q>1$, the approximation remains surprisingly good.

The agreement between the Monte Carlo estimate of positional information with its approximation based on positional error, observed in a controlled setting of Fig~\ref{f15}, is reflected in the analogous comparison on the real data highlighted in Fig~\ref{f16}. Here the Monte Carlo estimate is by about 0.1 bits larger than the estimate from positional error, corresponding to  a relative difference of roughly $2.5\%$, and formally still within the error bars of both estimates. Both analyses suggest that the gap gene quadruplet truly is in the small noise regime (note that this might not be true for single genes or gene pairs), and that tractable approximations to positional information are therefore available. 

%
%
%  DISCUSSION
%
%
%
%
%
%
\section{Discussion}

To generate a  differentiated body plan during the development of a multicellular organism, cells with identical genetic material need to reproducibly acquire distinct cell fates depending on their position in the embryo. The mechanisms of establishment and acquisition of such ``positional information'' have been widely studied, but the concept of positional information itself has, surprisingly, eluded  formal definition. Here we have provided a mathematical framework for  positional information and positional error based on information theory. These are principled measures for quantifying how much knowledge cells can gain about their absolute location in the embryo---and thus how precisely they can commit to correct cell fates---by locally reading out noisy gene expression profiles of (possibly multiple) morphogen gradients. From this broad perspective, our framework is a mathematical realization of the classic ideas put forth by Wolpert almost fifty years ago \cite{Wolpert69}.

To illustrate such an approach, we applied the framework of mutual information to the system of the four major gap genes (Knirps, Kr\"uppel, Giant, and Hunchback) that carry information about the nuclear positions of the central $80\%$ of the major axis of the \emph{Drosophila} embryo, a short report of which has been published recently \cite{PNASPI}. Our goal here was to provide a detailed account of the corresponding mathematical developments with a focus on the technical aspects of the analyses, which we extended to facilitate wider applicability to other patterning systems. Lastly, we are reporting on previously unpublished datasets (B, C, D) in the gap gene system that demonstrate consistency of our estimates across independently performed experiments.

An information-theoretic formulation of positional information has a number of very attractive features. First, it is mechanism independent. Many plausible mechanisms exist for the establishment of spatial gene expression patterns, involving the processes of gene regulation, signaling, diffusion, controlled degradation etc. However, the only quantities that matters for positional information are the shapes and co-variabilities of the expression profiles. This is  particularly intriguing as the mechanisms \emph{establishing} the gene expression patterns should naturally involve spatial coupling. For example, gene products diffuse across several nuclear distances in the early fly embryo \cite{Little:2013}. The spatial aspect of the problem would seemingly suggest that our definition of positional information which considers gene expression locally---separately at each spatial location $x$---is somehow insufficient or incomplete. However, irrespective of the type of non-local spatial coupling, what ultimately matters for positional information is the final pattern at the local scale, where nuclei make decisions, which is consistent with Wolpert's original idea. Conceptually we have provided a clear distinction between the mechanistic processes that generate expression profiles and the positional information that these profiles carry, and we hope that this separation will positively contribute to subsequent discussions in the field.

The second feature is that our definition of positional information is a priori free of assumptions of what specific geometric features of the expression profiles---boundaries, domains, slopes, etc.---``carry'' or encode the information. Much prior work has focused on the sharpness of an expression boundary as a proxy for positional information \cite{Meinhardt:1983,Crauk:2005,Dahmann:2011,Zhang:2012,Lopes:2012}. It is unclear, however, how to generalize this approach to systems with more than one expression boundary, and, even more profoundly, we show that the boundary sharpness is \emph{not} the feature that actually maximizes positional information. In contrast, information-theoretic framework makes it clear in what particular way profile shapes and their (co-)variabilities need to be combined into an appropriate measure of positional information. This measure, as well as the associated positional error, easily generalize to patterning systems with an arbitrary number of gene gradients.

Lastly, the third feature is that positional information inherits all the attractive features of mutual information. The numerical value of mutual information can be interpreted in terms of  the number of distinguishable states, or, in the developmental context, as the number of distinguishable positions and cell fates.  Mutual information is reparametrization invariant, that is, its value does not depend on what units, or what scale, e.g., log vs linear, the expression levels are measured on. To estimate the information, carefully worked out estimation procedures exist and can be adapted to the developmental context. Lastly, as additional experimental variability can only decrease the information, the computed values will always be conservative lower bound estimates of the true information; that is, as we learn about the experimental sources of variability (and remove them, either by designing better experiments or improving data processing / normalization), we should be getting increasing estimates for positional information that approach the true, biologically relevant, value. 

Methodologically, we have hopefully provided enough detail to make this framework applicable to different systems and experimental setups. In particular, we have shown how data can be aligned and normalized if necessary, how different partial stainings can be  merged into a consistent dataset, and how analysis methods, including inference from data, numerical computation, and information estimation, should be performed on real experimental data. Importantly, we have proposed how information can be estimated in the small noise regime (which is likely applicable in different systems), and how the consistency of these estimates can be validated.

A major shortcoming of the proposed analysis framework is in assuming that positional information is represented in steady-state expression patterns in a particular time window. In the gap gene system, the expression patterns are dynamic on the time scale of nuclear cycles. While they appear most stable in the nuclear cycle 14 and in the particular time class we chose for our analysis, whether or not that constitutes a true steady state has been debated \cite{bergman}. There exist other systems, e.g., the segmentation clock in vertebrate somite formation, which are intrinsically dynamic. While it is possible that positional information in all of these systems is encoded in the expression levels in particular time windows, it is (at least in principle) also possible that the positional information is encoded in full gene expression \emph{trajectories}, i.e., temporal sequences of gene expression levels. Extending the information-theoretic approach presented here to include the temporal aspect is an important future research direction.

The application of our methods to the \emph{Drosophila} gap gene system has generated several results beyond those reported in Ref~\cite{PNASPI} which we would like to highlight. First, the results are extremely consistent across 4 different datasets, with fractional std in information estimates of below 5\%. This scatter is comparable to the estimation errors on single datasets, indicating the extremely high biological reproducibility in the observed system, as well as very stringent control over the experiment and data analysis. Second, the analysis of positional error shows that positional information is encoded in the parts of the expression profile that have high slopes (spatial derivatives), further weakening the interpretation of gap genes as providing sharp boundaries between expression domains. Third, we find that at the level of gene quadruplet, the information is represented very redundantly. While beyond the scope of this paper, this suggests a very attractive interpretation where this redundancy could be used for robustness to different external perturbations, by allowing a measure of ``error correction'' in the down stream layer \cite{Gierer91}. For example, particular gap gene readout mechanisms could respond by choosing correct cell fate assignments not only to wild-type gap gene expression patterns, but also to certain other patterns generated by genetic or environmental perturbations. Finally, the difference between information estimates with and without {\tt X} alignment implies that a noticeable fraction of biological variability across the embryos consists of rigid shifts of the full gap gene expression pattern along the AP axis. This could suggest that the biological system cares less about the absolute position of each nucleus in the embryo, and more about their relative positions (i.e., order along the AP axis). In other words, certain types of correlated positional errors---e.g., where all nuclei collectively shift left or right---would be less harmful than others, because the in one scenario the ordering of the cell fates would be preserved, while in the other it would be randomly reshuffled. This is a topic of our future research.

A major shortcoming of the current experiments is their restriction to extracting expression profiles from below the dorsal embryo surface. The resulting analyses---including this one---therefore confound the expression levels of neighboring nuclei with the levels in the interstitial cytoplasmic space, while biologically meaningful levels are only the nuclear ones. Ideally a data set would contain expression levels on a nucleus-by-nucleus basis \cite{Myasnikova:2001,Surkova:2008,Myasnikova:2009}, possibly encompassing all nuclei in 3D reconstructed embryos \cite{Fowlkes:2008}. We  stress, however, that all our analysis methods presented here can be easily extended and applied to data sets containing discrete nuclear expression levels.  The emphasis here and in the associated experimental paper \cite{MSB} was on a careful assessment of systematic measurement errors and on achieving a throughput of $N\le100$ embryos per data set, necessary for computing the covariance across genes and embryos. The current experimental setups were feasible for these goals, but this would have to be reassessed for nuclear and 3D data sets.

\end{document}